\documentclass{imamat}
\gridframe{N}
\usepackage{modnatbib}
\usepackage{latexsym,amssymb,lastpage} 
\usepackage{graphicx,amsfonts}
\usepackage{times,mathptmx,bm,amsmath,amsthm}
\usepackage{dcolumn}
\usepackage{hyperref}% add hypertext capabilities
\usepackage{xcolor}
\usepackage{tikz}
\usepackage{ulem}
\usepackage{pstricks}
 \newtheoremstyle{theorem}{6pt}{6pt}{\rm}{}{\sffamily}{ }{ }{}
 \theoremstyle{theorem}

  \newtheoremstyle{thm}{6pt}{6pt}{\rm}{}{\sffamily}{ }{ }{}
 \theoremstyle{thm}

 \newtheoremstyle{lemma}{6pt}{6pt}{\rm}{}{\sffamily}{ }{ }{}
 \theoremstyle{lemma}
 
 \newtheoremstyle{lem}{6pt}{6pt}{\rm}{}{\sffamily}{ }{ }{}
 \theoremstyle{lem}

\newtheoremstyle{case}{6pt}{6pt}{\rm}{}{}{. }{ }{}
 \theoremstyle{case}

 \newtheoremstyle{statement}{6pt}{6pt}{\rm}{}{\sffamily}{ }{ }{}
\theoremstyle{statement}

 \newtheoremstyle{corollary}{6pt}{6pt}{\rm}{}{\sffamily}{ }{ }{}
 \theoremstyle{corollary}

  \newtheoremstyle{defi}{6pt}{6pt}{\rm}{}{\sffamily}{ }{ }{}
 \theoremstyle{defi}

  \newtheoremstyle{cor}{6pt}{6pt}{\rm}{}{\sffamily}{ }{ }{}
 \theoremstyle{cor}

\newtheoremstyle{example}{6pt}{6pt}{\rm}{}{\sffamily}{ }{ }{}
\theoremstyle{example}

\newtheoremstyle{remark}{6pt}{6pt}{\rm}{}{\sffamily}{ }{ }{}
\theoremstyle{remark}

\newtheoremstyle{approximation}{6pt}{6pt}{\rm}{}{\sffamily}{ }{ }{}
\theoremstyle{approximation}

\newtheoremstyle{scheme}{6pt}{6pt}{\rm}{}{\sffamily}{ }{ }{}
\theoremstyle{scheme}

\newtheoremstyle{Algorithm}{6pt}{6pt}{\rm}{}{\sffamily}{ }{ }{}
\theoremstyle{Algorithm}

 \newtheoremstyle{Remark}{6pt}{6pt}{\rm}{}{\sffamily}{ }{ }{}
 \theoremstyle{Remark}

\newtheoremstyle{Lemma}{6pt}{6pt}{\rm}{}{\sffamily}{ }{ }{}
\theoremstyle{Lemma}

\newtheoremstyle{Assumption}{6pt}{6pt}{\rm}{}{\sffamily}{ }{ }{}
\theoremstyle{Assumption}

\newtheoremstyle{Proposition}{6pt}{6pt}{\rm}{}{\sffamily}{ }{ }{}
\theoremstyle{Proposition}

\newtheoremstyle{prop}{6pt}{6pt}{\rm}{}{\sffamily}{ }{ }{}
\theoremstyle{prop}

\newtheoremstyle{rem}{6pt}{6pt}{\rm}{}{\sffamily}{ }{ }{}
 \theoremstyle{rem}

\newtheoremstyle{hypo}{6pt}{6pt}{\rm}{}{\sffamily}{ }{ }{}
 \theoremstyle{hypo}

  \newtheoremstyle{Step}{6pt}{6pt}{\rm}{}{}{ }{ }{}
 \theoremstyle{Step}

 \newtheoremstyle{lema}{6pt}{6pt}{\rm}{}{\sffamily}{ }{ }{}
 \theoremstyle{lema}
 
\newcommand{\bra}[1]{\left(#1\right)} \newcommand{\Bra}[1]{\left[#1\right]}

\newcommand{\ellc}{\ell_{\text{T}}}
\newcommand{\kc}{k_{\text{T}}}
\newcommand{\rT}{\rho_{\text{T}}}
\newcommand{\rH}{\rho_{\text{H}}}
\newcommand{\rSN}{\rho_{\text{SN}}}
\newcommand{\rEP}{\rho_{\text{EP}}}
\newcommand{\rBD}{\rho_{\text{BD}}}
\newcommand{\vP}{{\bf P}}

\begin{document}
	%%%%%%%%%%%%%%% this is title page style%%%%%%%%%
	\title{Stationary peaks in a multivariable reaction--diffusion system: \\ Foliated snaking due to subcritical Turing instability}
	
	\author{{\sc Edgar Knobloch$^1$ and Arik Yochelis$^{2,3}$}\\[2pt]
		$^1$Department of Physics, University of California, Berkeley, California 94720, USA
		\\[2pt]
		$^2$Department of Solar Energy and Environmental Physics, Swiss Institute for Dryland Environmental and Energy Research, Blaustein Institutes for Desert Research, Ben-Gurion University of the Negev, Sede Boqer Campus, \\ Midreshet Ben-Gurion 8499000, Israel
		\\[2pt]
		$^3$Department of Physics, Ben-Gurion University of the Negev, \\ Be'er Sheva 8410501, Israel
		\\[6pt]
		{\rm [Received on \today]}}
	\pagestyle{headings}
	\markboth{E. Knobloch and A. Yochelis}{\rm Stationary peaks in a multivariable reaction--diffusion system}
	\maketitle
	%%%%%%%%%%%%%%%%%
	\begin{abstract}
		{An activator-inhibitor-substrate model of side-branching used in the context of pulmonary vascular and lung development is considered on the supposition that spatially localized concentrations of the activator trigger local side-branching. The model consists of four coupled reaction-diffusion equations and its steady localized solutions therefore obey an eight-dimensional spatial dynamical system in one dimension (1D). Stationary localized structures within the model are found to be associated with a subcritical Turing instability and organized within a distinct type of foliated snaking bifurcation structure. This behavior is in turn associated with the presence of an exchange point in parameter space at which the complex leading spatial eigenvalues of the uniform concentration state are overtaken by a pair of real eigenvalues; this point plays the role of a Belyakov-Devaney point in this system. The primary foliated snaking structure consists of periodic spike or peak trains with $N$ identical equidistant peaks, $N=1,2,\dots \,$, together with cross-links consisting of nonidentical, nonequidistant peaks. The structure is complicated by a multitude of multipulse states, some of which are also computed, and spans the parameter range from the primary Turing bifurcation all the way to the fold of the $N=1$ state. These states form a complex template from which localized physical structures develop in the transverse direction in 2D.}
		{localized states, homoclinic snaking, wavenumber selection, reaction--diffusion systems}
	\end{abstract}
	%%%%%%%%%%%%%%%%%%%%%%%%%%%%%%%
	
	%%%%%%%%%%%%%%section A%%%%%%%%%
	\section{Introduction}
	%%%%%%%%%%%%%%%%%%%%%%%%%%%%%
	
	Reaction--diffusion (RD) type models are widely studied as a laboratory for the study of pattern formation in spatially extended systems that are driven far from equilibrium~\cite{ch93,CG09}, as exemplified by physico-chemical autocatalytic systems~\cite{ertl-i95,epstein1998introduction,pismen06,kapral2012chemical}, biological and medical applications~\cite{maini2001mathematical,murray2007mathematical,ksp98,ksp08}, models of vegetation cover in ecology~\cite{meron2015book} and even nonlinear optics~\cite{arecchi1999pattern}. Even though RD models cannot capture the complexity of many real-world applications, especially in the medical context, they share many generic similarities with them and so are often exploited to provide a qualitative understanding of the mechanistic basis of pattern formation and evolution in a wide variety of natural settings, including periodic pigmentation on animal skins, intracellular $\text{Ca}^{2+}$ and actin waves, electrical impulses in neurons and in the heart, swarming phenomena in bacterial colonies or flocks of birds and fish, spatially localized resonances in the inner ear, and the development of tissues and organs. In fact, this approach has already been outlined in the seminal works of Turing on morphogenesis~\cite{tu52} and of Hodgkin and Huxley on action potentials in the giant squid axon~\cite{hodgkin1952quantitative}. Since then several prototypical RD models, such as the FitzHugh--Nagumo~\cite{fitzhugh1961impulses,nagumo1962active}, the Keller--Segel~\cite{keller1971model} and the Gierer--Meinhardt~\cite{gm72} equations, have been subjected to detailed study in order to gain further insight into many of these systems.
	
	In the biomedical context, RD models are often referred to as activator--inhibitor (AI) systems, where the activator is locally expressed on a fast timescale but diffuses slowly while the inhibitor diffuses rapidly. These two kinetic ingredients are essential for the presence of the Turing instability~\cite{tu52} that gives rise not only to periodic patterns~\cite{nagorcka1983evidence,castets1990experimental,ouyang1991transition,lengyel1992chemical,kondo2002reaction,garfinkel2004pattern,maini2006turing,lin2009spots}, such as striped or hexagonal structures, but also to spatially localized structures that may range from stationary~\cite{yochelis2008formation,brena2014subcritical,purwins2010dissipative} or propagating~\cite{elphick1990patterns,yochelis2008generation} single peaks to groups of peaks. While a complete mathematical framework for studying these structures has not yet been established, uncovering partial mechanisms behind biomedical pattern-formation phenomena remains of utmost importance for understanding functional aspects and application design behind drug delivery systems or implants. On the other hand, mechanistic studies of medical systems are a fertile source of new mathematical questions, especially in the context of model simplification and applicability as a useful description of highly detailed and multiscale phenomena. In the present work, our interest lies in spatially localized states that are symmetric under spatial reflections $x\to-x$ in one dimension (a symmetry broken by propagating excitable pulses) and their apparent role in initiating side-branching in mammalian organ development~\cite{caduff1986scanning,roth2005neonatal,warburton2008order,metzger2008branching,varner2017computational,hannezo2019multiscale}.
	
	\subsection{Localized states and homoclinic snaking}
	
	Existence and stability of single steady-state localized states as well as groups of such states have been intensively studied in the last two decades in the context of many application-driven model equations including nonlinear optics~\cite{parra2016dark}, Faraday-waves~\cite{richter2005two,yochelis2006reciprocal,burke2008classification,dawes2008localized,alnahdi2014localized}, fluid convection~\cite{batiste2006spatially,jacono2011magnetohydrodynamic,beaume2013convectons}, solidification~\cite{archer2014solidification,thiele2019first}, electrically charged molecules~\cite{gavish2017spatially}, vegetation structures~\cite{zelnik2013regime,zelnik2015gradual,zelnik2016localized,zelnik2017desertification,zelnik2018implications}, hair-root formation in plants~\cite{brena2014mathematical,verschueren2017} and vascular mesenchymal cells~\cite{yochelis2008formation,yochelis2008front}. The mathematical organization of coexisting localized states in one space dimension (1D) is broadly attributed to the homoclinic snaking (HS) phenomenon~\cite{knobloch2015spatial}, the commonest manifestation of which is the so--called snakes--and--ladders structure of the snaking or pinning region~\cite{burke2007snakes}. Other types of organization, referred to as collapsed HS~\cite{burke2006localized,yochelis2006reciprocal,yochelis2008front}, defect-mediated HS~\cite{ma2010defect}, slanted HS~\cite{dawes2008localized}, and foliated HS~\cite{glasner2012characterising,ponedel2016forced,parra2018bifurcation} also arise. The latter is of particular relevance in this work. In particular, homoclinic snaking takes the form of a pair of intertwined branches of spatial oscillations with odd and even numbers of peaks, interconnected by rungs consisting of asymmetric solutions, all of which are embedded in a homogeneous background state with complex spatial eigenvalues. In contrast, in foliated snaking the basic structure is a single peak that coexists with separate branches of equispaced multipeak states. These branches are also interconnected by rung-like states but these consist of states with developing interpeak structures. This structure is associated with real spatial eigenvalues.
	
	%\begin{figure}[!t] %F1
	%		\centering\includegraphics[width=0.8\textwidth]{cartoon-eps-converted-to.pdf}
	%		\caption{Comparison between the classical snakes--and--ladders structure (a) and foliated snaking (b). The former consists of a pair of continuous intertwined branches connected by rung states consisiting of asymmetric states. The latter organizes equispaced peaks interconnected via rung states consisting of developing interpeak structures.}
	%		\label{fig:cartoon}
	%\end{figure}
	
	These different cases are distinguished not only by the nonlinear properties of the localized solutions in the parameter plane (e.g., the type of homoclinic and hetereclinic connection) but also by their origin, i.e., the nature of the primary bifurcation through which the localized solutions form. The origin can be attributed to physical symmetry-breaking characteristics of the system, for example, a subcritical Turing (or finite wavenumber) instability from which the snakes--and--ladders HS emerges~\cite{burke2007snakes}, while the organization of the resulting localized structures may be slanted (instead of being vertical) in parameter space owing to the presence of a conserved quantity or other nonlocal effects~\cite{dawes2008localized,thiele2013localized}. Although the factors that affect the HS properties may be obscured in specific model equations, the use of spatial dynamics can reveal both its origin and organization~\cite{champneys1998homoclinic,knobloch2015spatial}, an approach considerably advanced by Patrick Woods~\cite{woods1999heteroclinic,hunt2000cellular}.
	
	In the present work, we study in detail one such model system from this point of view. The model is of activator-inhibitor-substrate type and is motivated by increasing empirical evidence for the dominance of AI behavior as the driving mechanism behind branching~\cite{iber2013control,menshykau2019image} and specifically the development of the lungs~\cite{yao2007matrix}, although the details of this mechanism remain unclear owing to the multiscale nature of the processes involved that range from molecular to tissue level~\cite{varner2017computational,hannezo2019multiscale}.
	
	Beyond its potential applicability the model studied below also introduces new mathematical issues, arising from the fact that its spatial dynamics formulation is higher-dimensional than that of the prototypical models such as the Swift--Hohenberg or the complex Ginzburg--Landau equation or application-motivated models (e.g., Lugiato--Lefever, Gray--Scott and Gierer--Meinhardt models), all of which lead to a set of four first order ordinary differential equations (ODEs) for the spatial structure of the possible stationary states. It is this increased dimension that permits the appearance of new mathematical structures in this model. Despite this, we believe that the bifurcation structure we identify can also be found, with minor modifications, in these lower-dimensional models.
	
	\subsection{Localized states and side-branching}
	
	Meinhardt in 1976~\cite{meinhardt1976morphogenesis} proposed an activator-inhibitor-substrate approach for studying branching phenomena, which has since been employed in the context of pulmonary vascular and lung development~\cite{yao2007matrix}. The model is based on four fields $A$, $H$, $S$ and $Y$ that represent, respectively, the concentrations of an activator (BMP), an inhibitor (MGP), and the substrate (TGF-$\beta$/ALK1), as well as the concentration of an irreversible marker for differentiated endothelial cells~\cite{yao2007matrix}:
	\begin{subequations}\label{eq:AI}
		\begin{eqnarray}
			\frac{\partial A}{\partial t}&=&c\dfrac{SA^2}{H}-\mu A+\rho_{\text{A}} Y+D_{\text{A}} \nabla^2 A, \\
			\frac{\partial H}{\partial t}&=&cSA^2-\nu H+\rho_{\text{H}} Y+D_{\text{H}} \nabla^2 H, \\
			\frac{\partial S}{\partial t}&=&c_0-\gamma S-\varepsilon Y S+D_{\text{S}} \nabla^2 S, \\
			\frac{\partial Y}{\partial t}&=&d A-eY+\dfrac{Y^2}{1+fY^2}+D_{\text{Y}} \nabla^2 Y.
		\end{eqnarray}
	\end{subequations}
	Although $D_{\text{Y}}=0$ in the original formulation by Meinhardt (and in the studies that followed), here we take $D_{\text{Y}}$ as finite but much smaller than the smallest diffusion coefficient among all other fields, i.e, $D_{\text{Y}}\ll D_{\text{A}}$. This increased spatial order is inconsequential: what matters for us is that the $D_{\text{Y}}=0$ system is of sixth order in space. We also mention that although the number of parameters in system~\eqref{eq:AI} can be reduced, we follow here previous studies~\cite{yao2007matrix,yao2011matrix,guo2014mechanisms,xu2017turing} and do not do so. Additionally, to maintain fidelity to experiments with excess inhibitor (MGP)~\cite{yao2007matrix}, we employ the inhibitor control level, $\rho_{\text{H}}$, as a control parameter while keeping all other parameters fixed and within the range of previous studies: $c=0.002$, $\mu=0.16$, $\rho_{\text{A}}=0.005$, $\nu=0.04$, $c_0=0.02$, $\gamma=0.02$, $\varepsilon=0.1$, $d=0.008$, $e=0.1$, $f=10$, $D_{\text{A}}=0.001$, $D_{\text{H}}=0.02$, $D_{\text{S}}=0.01$, $D_{\text{Y}}=10^{-7}$.
	
\begin{figure}[!t] %F1
		\centering\includegraphics[width=0.8\textwidth]{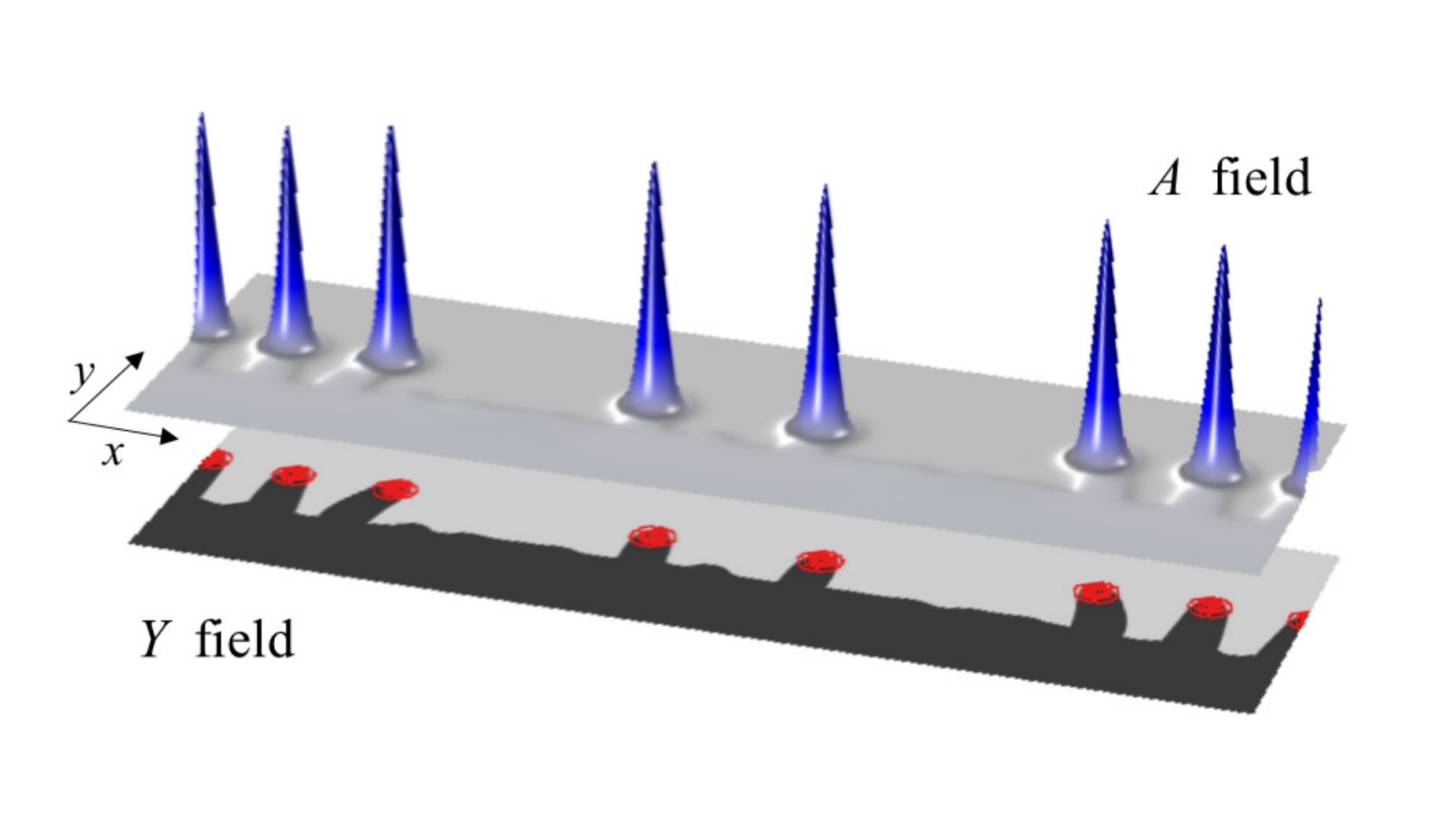}
		\caption{A representative snapshot of a solution obtained from a direct numerical simulation of Eqs.~\eqref{eq:AI} for $\rH=1.6\cdot 10^{-5}$ with periodic boundary conditions on $x \in [0,16]$ and Neumann boundary conditions on $y\in [0,5]$. The top panel represents the $A$ field while the bottom panel shows the $Y$ field, the dark colors indicating higher values of the field. The contour lines in the bottom panel mark the locations of the localized $A$ states shown in the top panel.}
		\label{fig:intro}
	\end{figure}

	Direct numerical simulations (DNS) of~\eqref{eq:AI} in two space dimensions (2D) show that side-branches develop via growing localized concentrations (top panel in Fig.~\ref{fig:intro}) that are locked to the edge of a propagating front generated by the differentiation field (bottom panel in Fig.~\ref{fig:intro}); the direction of propagation corresponds to the invasion of the light color region by the dark color region. Moreover, it has been shown through analysis of~\eqref{eq:AI} in 1D that the initiation of these side branches corresponds to the existence of localized solutions that bifurcate subcritically via a Turing instability from an expanding differentiated state uniform in the variable $x$ along the front~\cite{yochelis2020nonlinear}. In the following, we focus on the origin and properties of the localized structures shown in Fig.~\ref{fig:intro} and so formulate the problem in one space dimension, ignoring the motion of the front in the normal, $y$, direction. The latter will be the subject of a future publication.
	
	\subsection{Problem outline}
	
	We suppose that Eqs.~\eqref{eq:AI} are posed on the real line with periodic boundary conditions on the domain $x\in[0,L]$. The 1D domain and the choice of the boundary conditions mimic the planar interface that connects the $\vP_*$ and $\vP_0$ states prior to the formation of any peaks. Our aim is to reveal the mechanism responsible for the appearance of localized structures along this interface, prior to their expansion in the transverse direction. This picture is predicated on the assumption that these structures form on a faster timescale than that of the transverse growth, an assumption that will be tested in subsequent work on the 2D problem.
	
	In the following we describe the nature of the stationary nonuniform solutions focusing on the properties of spatially localized states when $L=60$. Our results are supplemented with additional results for other values of $L$. The steady solutions of this system solve an eight-dimensional system in space. Because of this higher dimension new behavior can arise that cannot occur in the four-dimensional systems hitherto studied. Thus our study of the system~\eqref{eq:AI} represents a nontrivial extension of existing work on spatially localized states.
	
	Specifically, we show that the steady localized states of \eqref{eq:AI} are organized in a bifurcation structure, which we call foliated snaking \cite{ponedel2016forced,parra2018bifurcation}. These states consist of trains of identical equispaced stationary peaks we refer to as primary states, as well as secondary states consisting of unequal amplitude peaks with unequal separations. We use the numerical continuation software AUTO~\cite{doedel2012auto} to continue both types of states as a function of the bifurcation parameter $\rH$. We show that these states are generated at small amplitude near the Turing bifurcation at $\rH=\rT$ creating a subcritical Turing pattern and then follow them through folds back to their termination near at finite amplitude, also near $\rT$. We examine the role of spatial resonances in generating the secondary states and identify the existence of an exchange point (EP), $\rH=\rEP$, which plays a role in the behavior of the system analogous to that of the Belyakov-Devaney (BD) point. We confirm our $L=60$ results by performing similar computations for $L=20$, 24 and 30, obtaining essentially identical results, and discuss the relation of finite domain numerical continuation results to a theoretical analysis of the unfolding of a global bifurcation at BD in Ref.~\cite{verschueren2018dissecting}. We conjecture that the foliated snaking structure we identify here is both robust and universal and should therefore be present in other systems of partial differential equations supporting peak-like states, ranging from branching in physiology to the formation of plant roots and of vegetation patches. In particular, we suggest that the structure we uncover arises generically in systems of two coupled spatially reversible reaction-diffusion equations on a line. The paper concludes with an outlook for future work.        
	
	\section{Linear analysis of uniform solutions}
	
\begin{figure}[!t] %F2
		(a)\centering\includegraphics[width=0.75\textwidth]{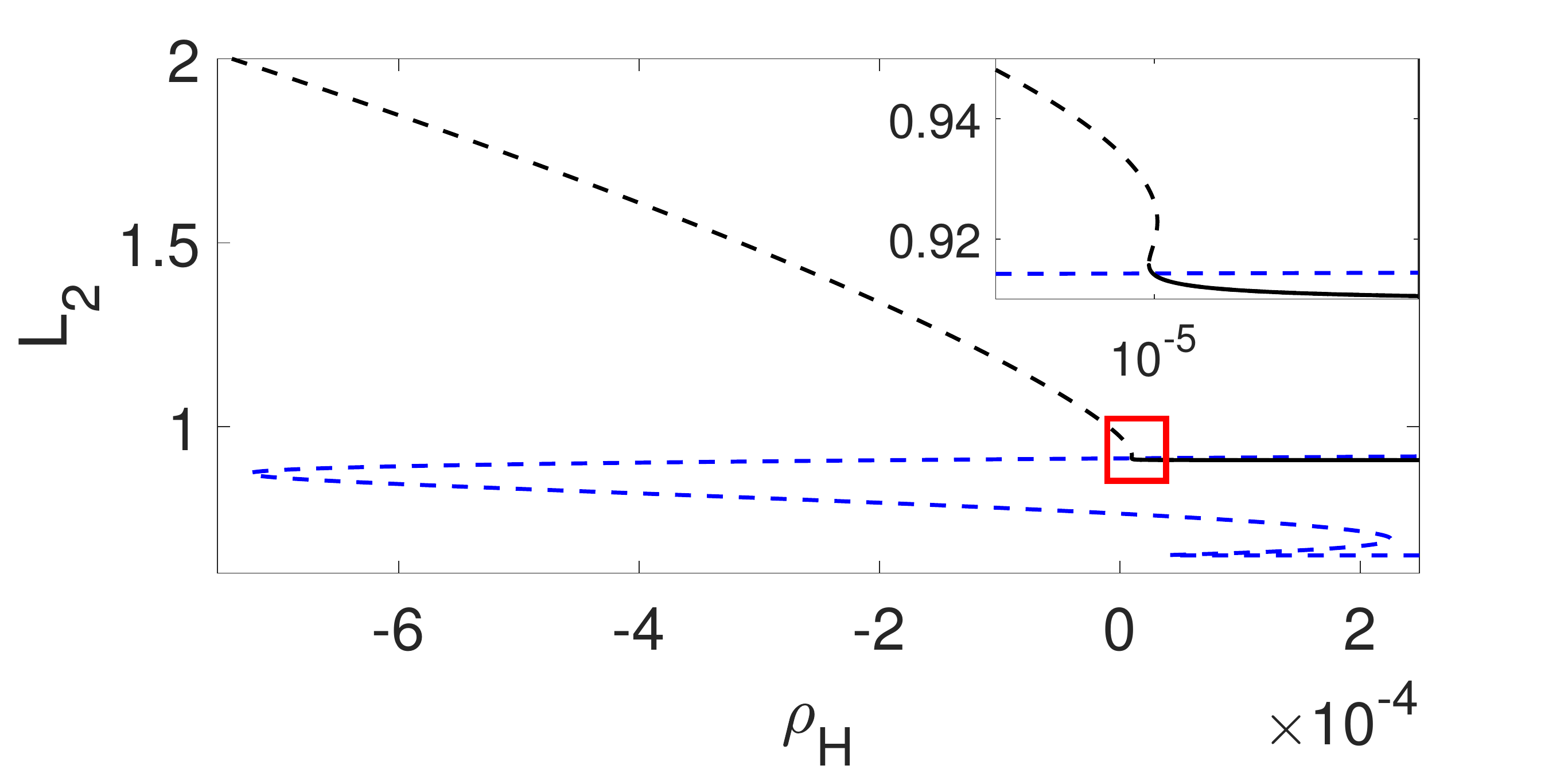} \vskip 0.2in
		(b)\centering\includegraphics[width=0.75\textwidth]{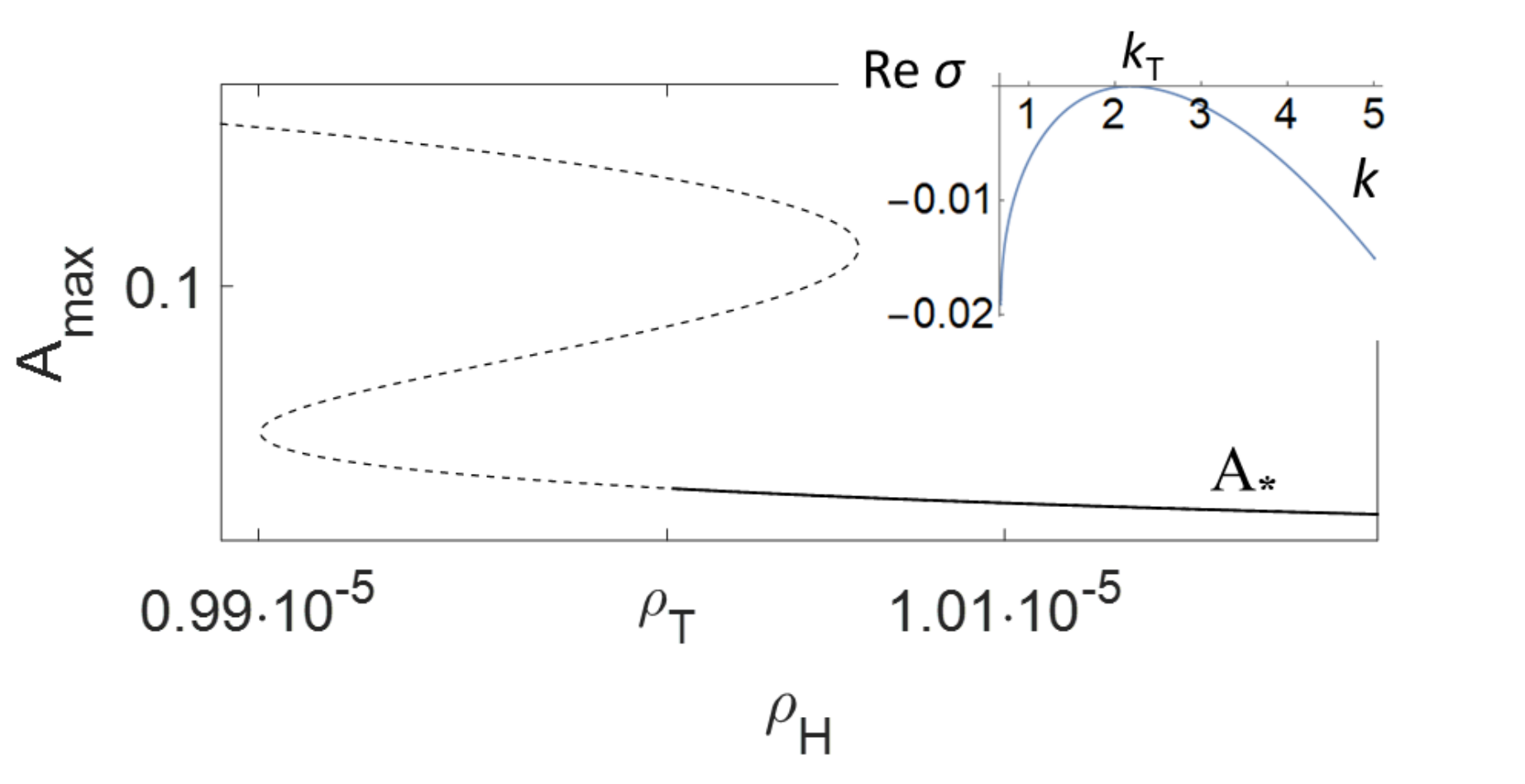}
		\caption{(a) Bifurcation diagram showing the coexistence and linear stability (solid lines) of uniform solutions of~\eqref{eq:AI} in terms of the ${\text L}_2$ norm (\ref{eq:norm}). The red rectangle marks the region of interest; there are two distinct families of uniform solutions in this region, a projection effect absent from panel (b). (b) Bifurcation diagram in terms of $A$ (within the region of interest), showing the onset of a Turing instability at $\rT \simeq 1.0011\cdot 10^{-5}$, with the inset depicting the dispersion relation, i.e., the real part of the growth rate $\sigma$ as a function of the wavenumber $k$ according to~\eqref{eq:turing}, where $\kc\simeq2.18$ is the critical wavenumber.}
		\label{fig:bif_uni}
	\end{figure}

	We begin by solving~\eqref{eq:AI} for steady spatially uniform states. In Fig.~\ref{fig:bif_uni}(a), we show that together with the `trivial' linearly stable solution $\vP_0\equiv (A_0,H_0,S_0,Y_0)=(0,0,c_0/\gamma,0)$ there are three or five additional nontrivial solutions that coexist for different values of $\rho_{\text{H}}$. Among these coexisting solutions, only one is linearly stable with respect to spatially uniform perturbations and we refer to this solution as $\vP_*\equiv (A_*,H_*,S_*,Y_*)$. Note that the front solutions that are shown in Fig.~\ref{fig:intro}, connect the $\vP_0$ and the $\vP_*$ states. The $\vP_*$ solution is thus the solution of interest in the present paper. 
	
	The homogeneous states $\vP_*$ are found on the bottom portion of a branch that resembles a backwards $S$, as shown in Fig.~\ref{fig:bif_uni}(b). The stability of the $\vP_*$ states is described by linearized equations about $\vP_*$,
	\begin{eqnarray}\label{eq:turing}
		\begin{pmatrix}
			A\\ 
			H\\ 
			S\\ 
			Y
		\end{pmatrix} - \begin{pmatrix}
			A_*\\ 
			H_*\\ 
			S_*\\ 
			Y_*
		\end{pmatrix} \propto e^{\sigma t + \lambda x}.
	\end{eqnarray}
	Thus $\sigma$ represents the growth rate in time of the perturbation while $\lambda$ represents its growth rate in space. For our spatial analysis we take $\sigma=0$ obtaining a quartic equation for $\lambda^2$, a consequence of the spatial reversibility of the system. 
	
\begin{figure}[!t] %F3
		\centering\includegraphics[width=0.75\textwidth]{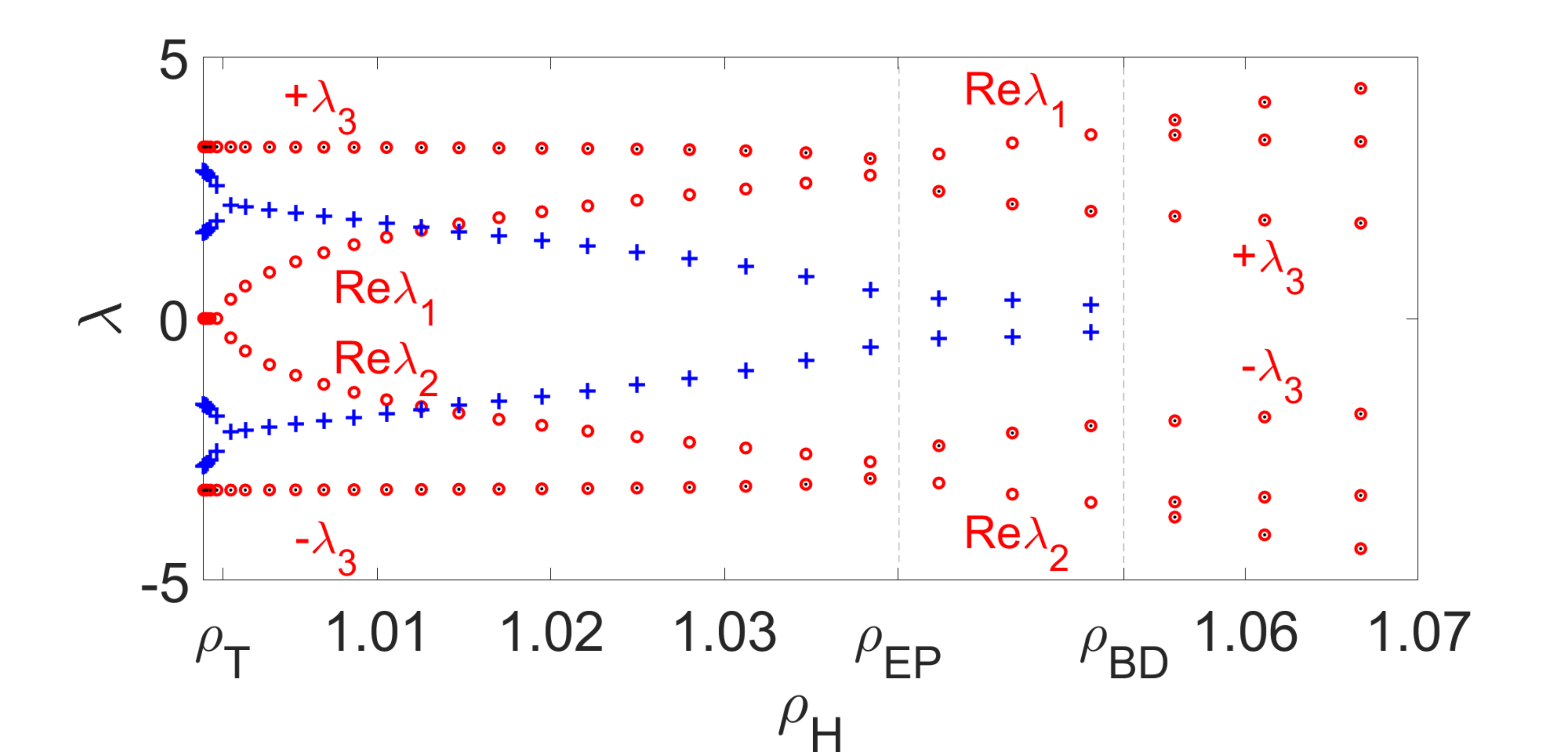}
		\caption{The six eigenvalues $\lambda$ as a function of $\rho_{\text{H}}$. Imaginary parts of $\lambda_{1,2}$ are indicated by the $+$ symbol. The point $\rBD\simeq 1.053\cdot 10^{-5}$ corresponds to the Belyakov-Devaney (BD) point where $\lambda_{1,2}$ become real, while $\rEP\simeq 1.04\cdot 10^{-5}$ indicates the exchange point (EP) where the eigenvalue $|\lambda_3|$ becomes smaller than $|{\rm Re}[\lambda_{1,2}]|$ as $\rH$ increases. The eigenvalues $\pm\lambda_4\sim \pm 800$ are not shown. The $\rH$ axis here and in all subsequent figures is scaled by $10^{5}$, unless stated otherwise.} \label{fig:eigs}
	\end{figure}

	In this case, because of the small value of $D_{\text{Y}}$, there is always a pair of large real eigenvalues $\pm \lambda_4$. On the lower branch to the right of the Turing bifurcation, $\rH>\rT$, there is a second pair of real eigenvalues, $\pm \lambda_3$ and a quartet of complex eigenvalues. At the Turing bifurcation $\rH=\rT\simeq 1.0011 \cdot 10^{-5}$, computed here for an infinite system (for $L=60$ the value is essentially identical), these complex eigenvalues collide pairwise on the imaginary axis, forming a pair of imaginary eigenvalues $\lambda_{1,2}=\pm i\kc$ of double multiplicity (see Fig.~\ref{fig:eigs}), where $\kc \simeq 2.18$ is the critical wavenumber at the Turing onset (see inset in Fig.~\ref{fig:bif_uni}(b)). For $\rH<\rT$ these split into four purely imaginary eigenvalues, a situation that corresponds to temporal instability of $\vP_*$, i.e., in this region ${\rm Re} [\sigma(k)]>0$ for a band of wavenumbers about $\kc$. With further decrease in $\rho_{\text{H}}$ the two imaginary eigenvalues closest to the origin collide at the origin, i.e., at $\lambda=0$, indicating the presence of the left fold of the S-shaped branch and above this fold the eigenvalues are real (not shown). On the middle part of the S-shaped branch there are three pairs of real eigenvalues and these become four pairs above the right fold.
	
	If we instead increase $\rH$ from $\rT$ we find that the imaginary parts of the eigenvalues $\pm\lambda_{1,2}$ gradually decrease, and that these eigenvalues become real at the Belyakov-Devaney (BD) point $\rBD\simeq 1.053\cdot 10^{-5}$ implying the disappearance of an intrinsic wavelength from the behavior of the system. Such a picture is familiar from similar behavior found, for example, in the Lugiato-Lefever equation in nonlinear optics when the homogeneous state is modulationally unstable \cite{parra2018bifurcation}. However, the steady states of this equation represent a four-dimensional spatial system. In the following we shall see that the `extra' eigenvalue $\lambda_3$ present in system \eqref{eq:AI} plays a new and significant role in the observed behavior. Figure \ref{fig:eigs} indicates the essential new feature arising from the presence of $\lambda_3$: as $\rH$ increases $|{\rm Re}(\lambda_{1,2})|$ increases and at $\rEP$ $|\lambda_3|$ falls below $|{\rm Re}(\lambda_{1,2})|$, i.e., the complex eigenvalues $\pm\lambda_{1,2}$ cease being leading eigenvalues and their role is taken by the real eigenvalues $\pm\lambda_3$, a fact that has a profound influence on the behavior of the system. This is because the leading eigenvalues generically determine the shape of localized structure near the background homogeneous state. In the following we refer to the transition at $\rEP\simeq 1.04\cdot 10^{-5}$ as an exchange point (EP). Figure \ref{fig:eigs} shows that this point is associated with a spatial {\it resonance}.
	
	We mention that a Turing or modulational instability is not a prerequisite for the presence of a BD point, although in cases in which a Turing instability is absent the localized structures and the associated foliated snaking can be traced to other bifurcations, such as a fold \cite{parra2018bifurcation} or a transcritical bifurcation \cite{ruiz}. However, the BD point continues to play a similar role to that here whenever it is present. In higher dimensions the EP point is expected to do the same.
	
	\section{Nonlinear results: A new type of foliated snaking}
	
	The Turing bifurcation at $\rH=\rT\simeq 1.0011 \cdot 10^{-5}$ provides a key to much of the nonlinear behavior observed in this system since it gives rise to a subcritical branch of periodic states with wavenumber $\kc$ and corresponding wavelength $\ellc\equiv 2\pi/\kc\simeq 2.88$. These periodic states, which extend towards $\rH>\rT$, are therefore temporally unstable, see~\cite{yochelis2020nonlinear} for details.
	
	Our interest is in the organization of the stationary localized states associated with this bifurcation. For this purpose, we employ numerical continuation~\cite{doedel2012auto} and compute solutions of the system~\eqref{eq:AI} written as a set of eight first order ODEs in space:
	\begin{subequations}\label{eq:AIode}
		\begin{eqnarray}
			A'&=&-a, \\
			a'&=&D^{-1}_{\text{A}} \bra{c\dfrac{SA^2}{H}-\mu A+\rho_{\text{A}} Y}, \\
			H'&=&-h, \\
			h'&=&D^{-1}_{\text{H}} \bra{cSA^2-\nu H+\rho_{\text{H}} Y}, \\
			S'&=&-s, \\ 
			s'&=&D^{-1}_{\text{S}} \bra{c_0-\gamma S-\varepsilon Y S}, \\ 
			Y'&=&-z, \\
			z'&=&D^{-1}_{\text{Y}} \bra{d A-eY+\dfrac{Y^2}{1+fY^2}}.
		\end{eqnarray}
	\end{subequations}
	Here the prime indicates differentiation with respect to $x$. Our results are summarized in the next section. These resemble the phenomenon of foliated snaking originally described in \cite{ponedel2016forced} in the context of a spatially forced two-variable PDE (or four-variable spatial ODE), and subsequently identified in the homogeneously forced Lugiato-Lefever equation in Ref.~\cite{parra2018bifurcation}; see also \cite{glasner2012characterising}. However, the present system is higher-dimensional and this fact is reflected in the observed behavior as discussed further below.

	\subsection{Localized states on large domains}
	
\begin{figure}[!t] %F4
		(a)\centering\includegraphics[width=0.75\textwidth]{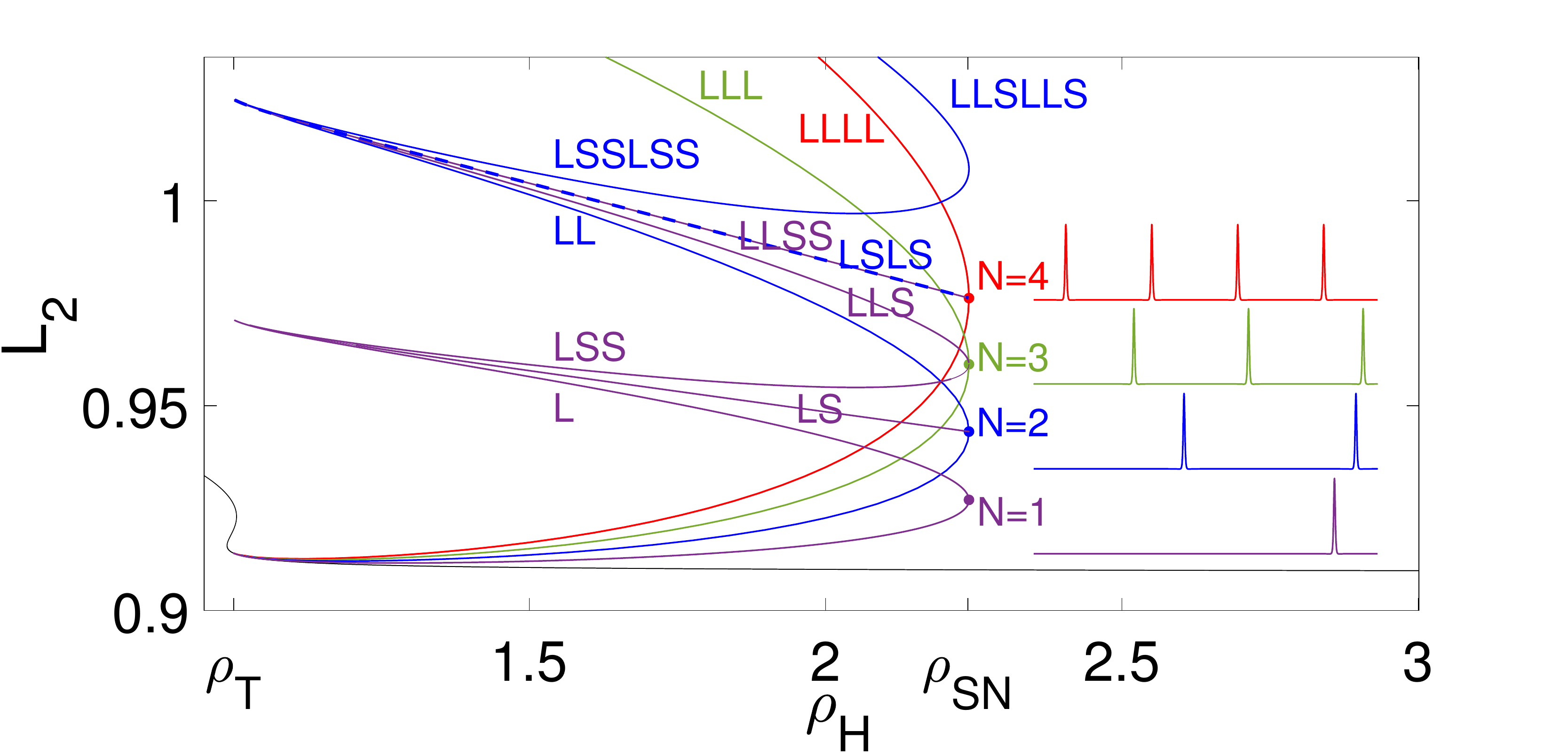}
		(b)\centering\includegraphics[width=0.75\textwidth]{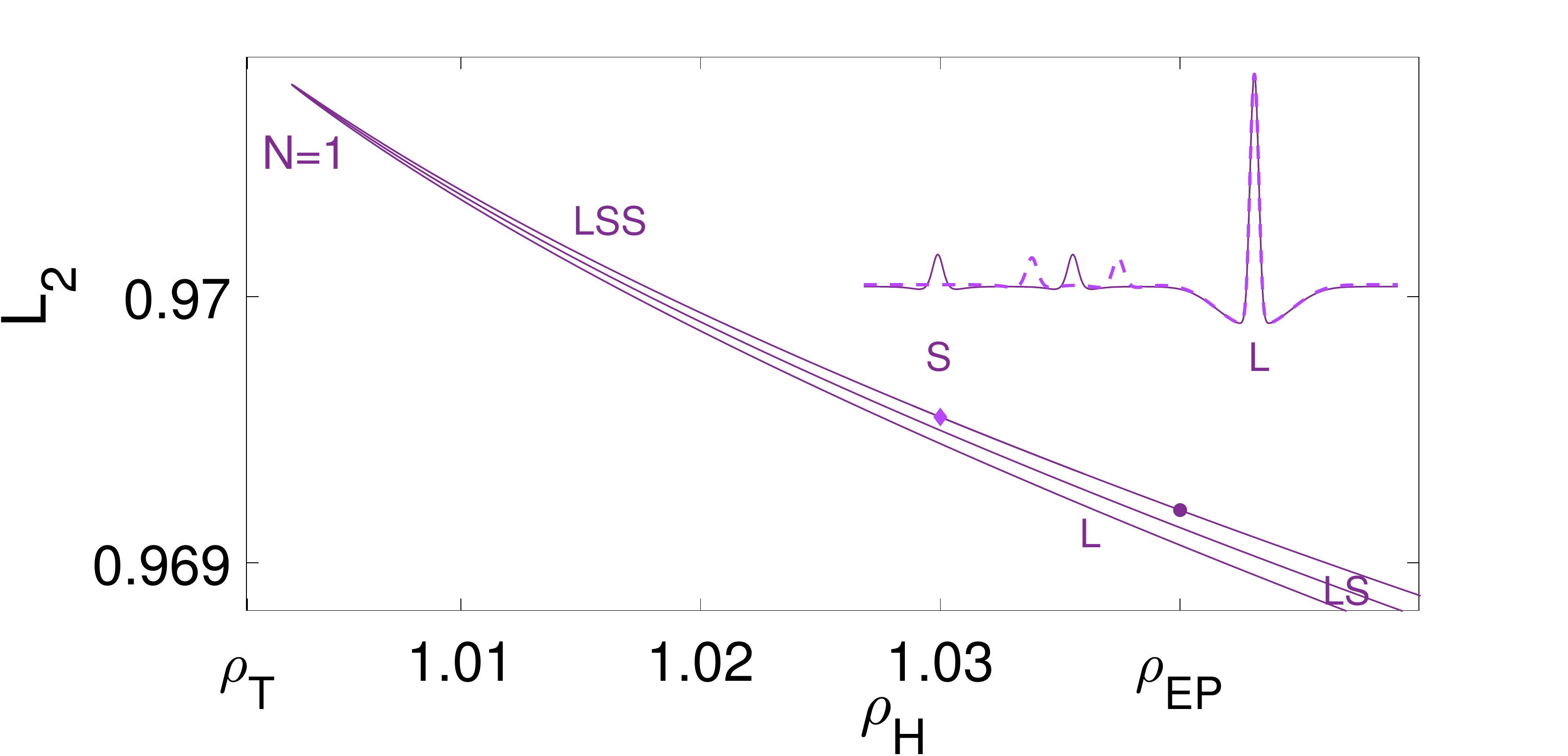}
		(c)\centering\includegraphics[width=0.75\textwidth]{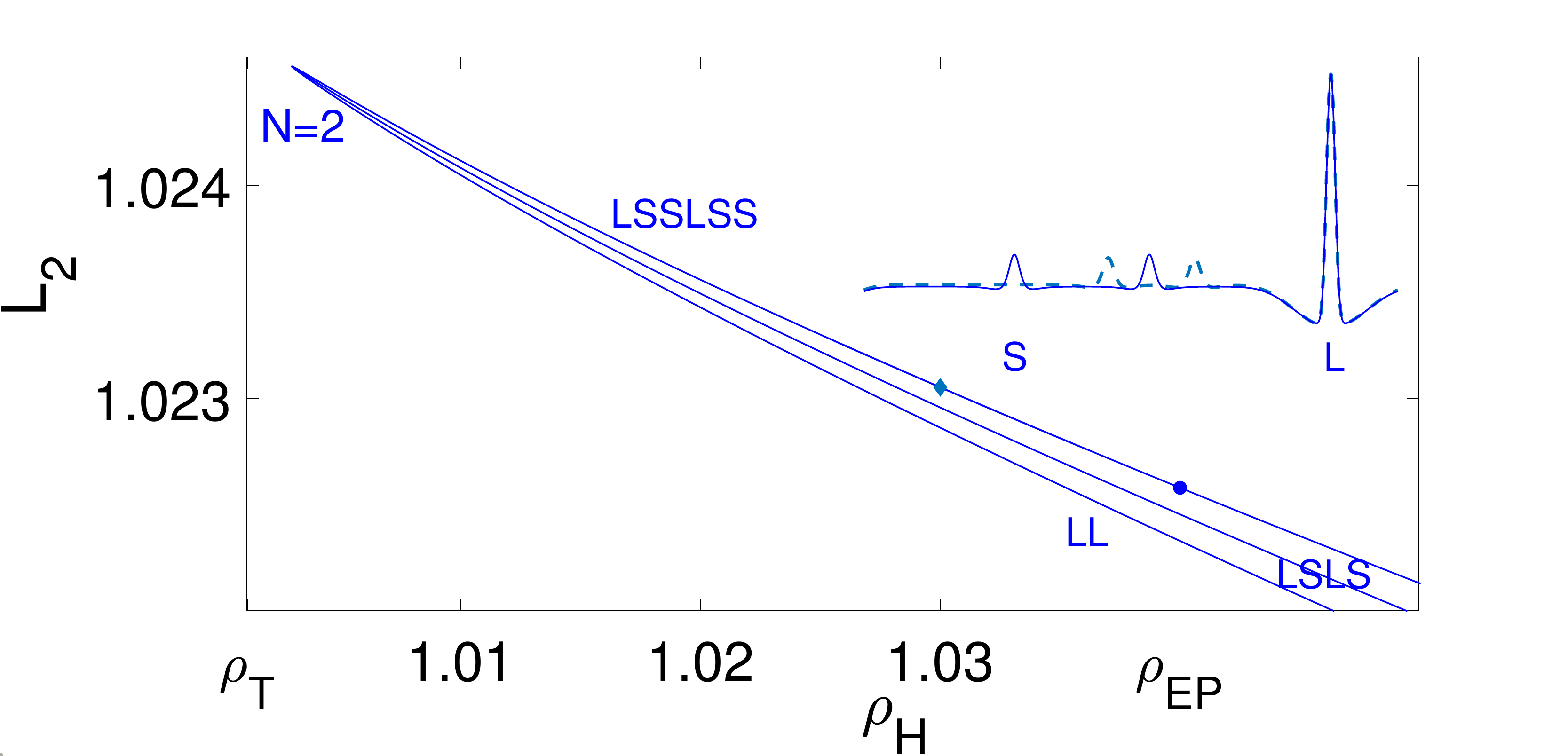}
		\caption{(a) Bifurcation diagram showing the primary $N=1,2,3,4$ branches originating near $\rT$ when $L=60$ as well as additional interconnecting branches. Color-coded profiles at locations indicated by dots are provided alongside and represent periodic states on the real line. (b,c) Detail of (a) showing that the finite amplitude termination points on the upper left of the (b) $N=1$ branch and (c) $N=2$ branch. In each case three branches are shown, labeled by the number of small (S) and large (L) peaks in the profile, with insets showing the profiles along LSS and LSSLSS at locations indicated by $\bullet$ (solid profiles) and $\blacklozenge$ (dashed profiles). The LSLS branch in (a) is shown dashed since it has almost identical ${\text L}_2$ norm to LLSS; branches of LLLS and LSSS are omitted (see Fig.~\ref{fig:profs4}(a)).} \label{fig:snaking}
	\end{figure}

\begin{figure}[!t] %F5
		(a)\centering\includegraphics[width=0.75\textwidth]{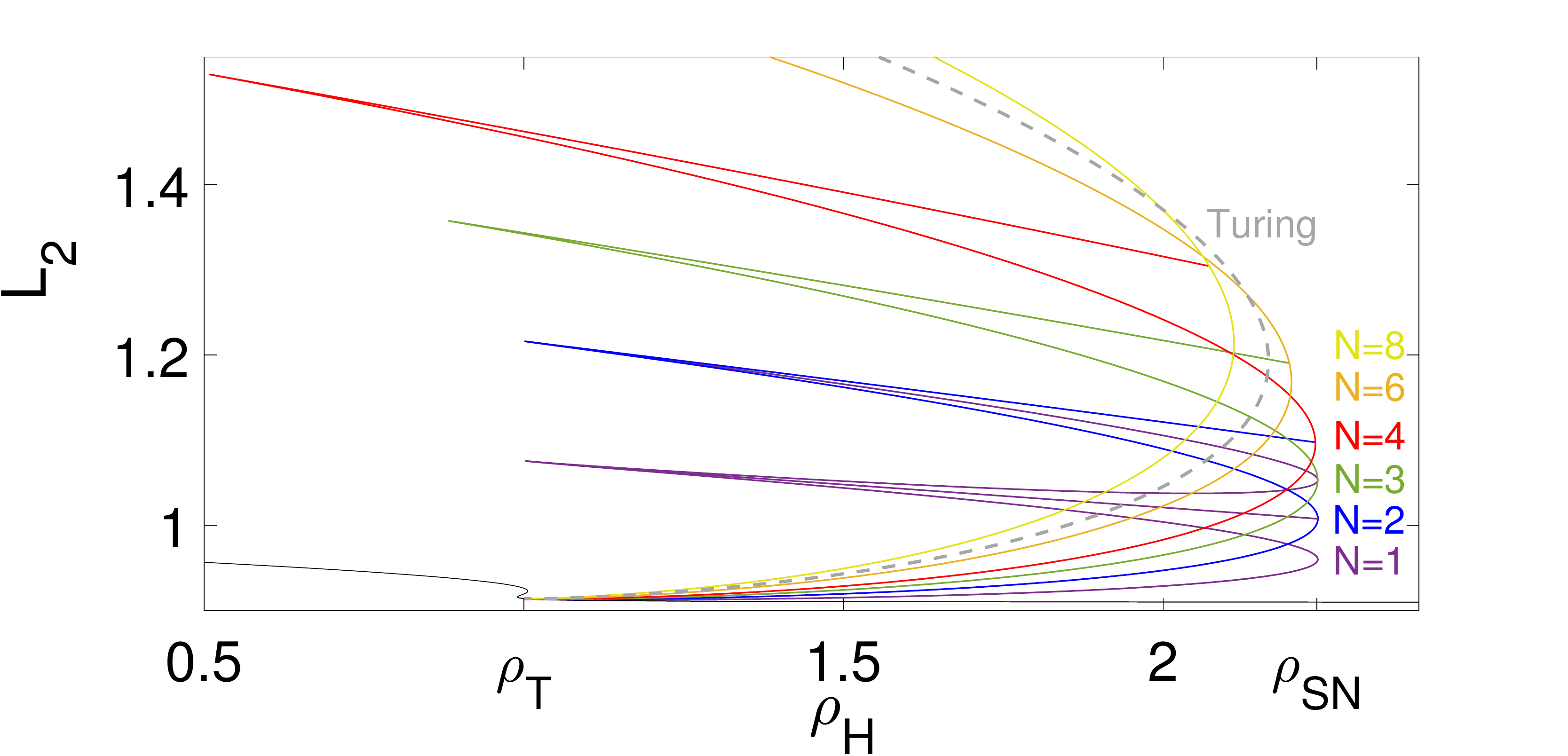}
		(b)\centering\includegraphics[width=0.75\textwidth]{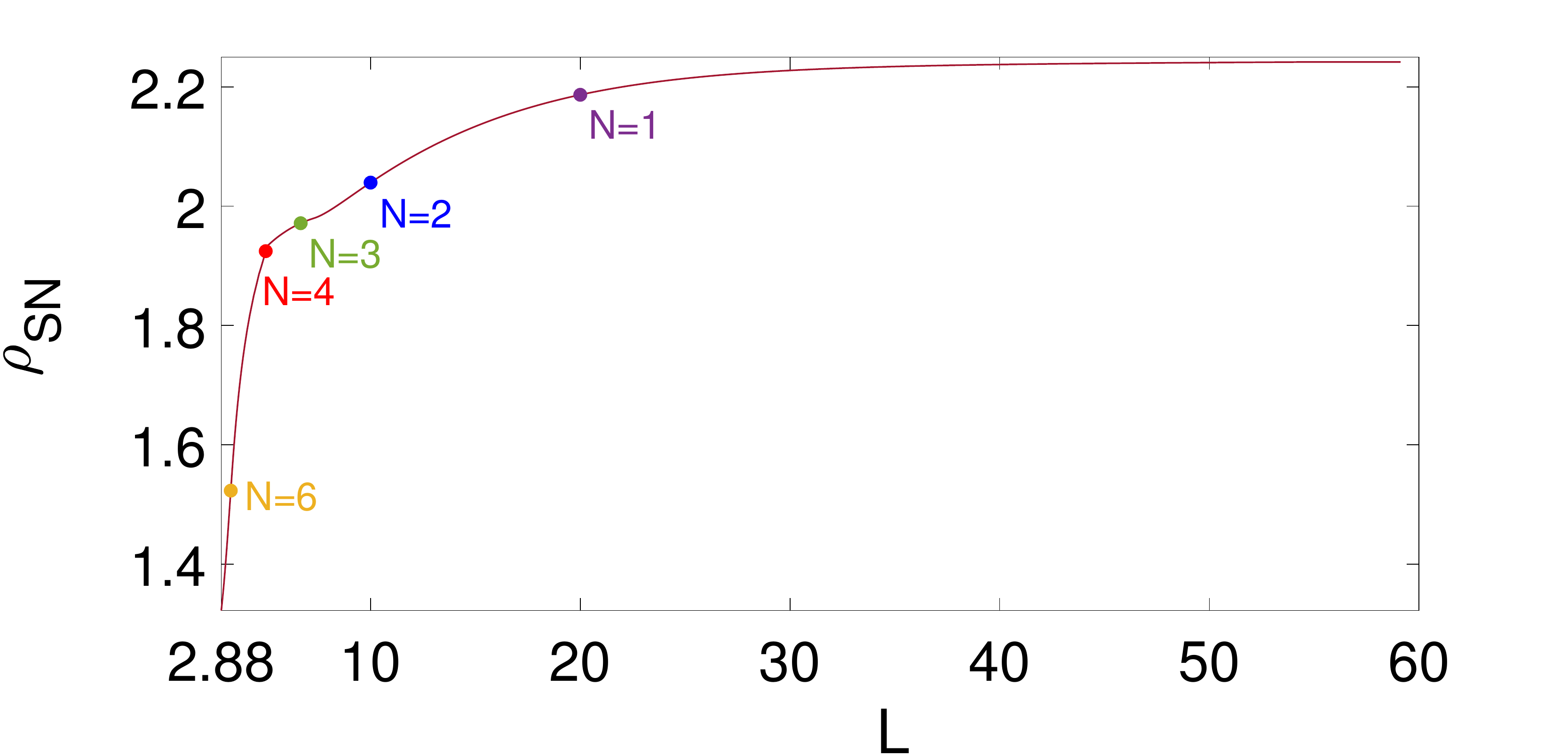}
		\caption{(a) Bifurcation diagram for $L=20$ showing that the behavior of the system departs from its universal behavior (Fig.~\ref{fig:snaking}(a)) when the number $N$ of peaks within the domain becomes too large. (b) Location $\rSN$ of the right folds as a function of the domain size $L$, obtained by continuation in $L$ of the location of the right fold of the $N=1$ primary state. This fold is always closest to the accumulation point of the different $N$-peak folds. The figure shows that $\rSN$ is close to its asymptotic value for $L\gtrsim 30$.}
		\label{fig:rfolds}
	\end{figure}

	In Fig.~\ref{fig:snaking}(a), we show the basic bifurcation diagram for a domain of length $L=60$, showing the branch of homogeneous states near $\rT$, as well as branches of $N=1$, 2, 3 and 4 identical equispaced peaks in the domain, as shown in the inset. The diagram shows the ${\text L}_2$ norm of the solutions, 
	\begin{equation}\label{eq:norm}
		\text{L}_2=\sqrt{L^{-1}\int_0^L \text{d}x\Bra{A^2+\bra{A'}^2+H^2+\bra{H'}^2+S^2+\bra{S'}^2+Y^2+\bra{Y'}^2}},
	\end{equation}
	as a function of the parameter $\rH$. Observe that on sufficiently large domains ($L\gg\ellc$) the low $N$ branches all bifurcate subcritically, and that all subsequently undergo a fold near essentially the same value $\rH=\rSN\simeq 2.241 \cdot 10^{-5}$ (see Fig.~\ref{fig:rfolds}), before turning around towards smaller values of $\rH$. In the following, we examine in detail the behavior of these primary branches near their point of origin near the Turing point at $\rT$, their folds on the right, near $\rSN$, and finally those on their left since all $N$-peak branches appear to terminate back at the Turing point $\rT$, although this time at finite amplitude.
	
	In fact, Figs.~\ref{fig:snaking}(b,c) show that the primary branches do not terminate at $\rT$ (for $\rH<\rT$ the leading eigenvalues $\lambda$ are imaginary thereby preventing approach to $\vP_*$) but instead turn around in a narrow fold, and that in this region they start to develop small intervening peaks, i.e., in this region the solutions no longer consist of equispaced identical peaks.
	
	To understand this behavior qualitatively recall that in the absence of locking between adjacent peaks, i.e., beyond the so-called Belyakov-Devaney transition, $\rH>\rBD$, where all the spatial eigenvalues $\lambda$ are real, we expect all primary solutions to consist of equispaced peaks. In contrast, when $\rT<\rH<\rBD$ the spatial eigenvalues are complex and such eigenvalues imply the presence of oscillations in the tail of each peak and hence the presence of locking of adjacent peaks at distinct separations. In other words, we expect a large variety of states for $\rT<\rH<\rBD$ owing to the presence of spatial locking but a much simpler situation for $\rH>\rBD$ where locking is absent. This is in fact the picture that is consistent with the behavior observed in four-dimensional systems~\cite{parra2018bifurcation,verschueren2018dissecting}.
	
	The present system is effectively six-dimensional, however, if we ignore the pair of large real eigenvalues. Figure~\ref{fig:eigs} shows the behavior of the six spatial eigenvalues along the branch of homogeneous states as $\rH$ increases through $\rH=\rT$ for our parameter values. We see that for $\rT<\rH<\rEP<\rBD$ the complex eigenvalues have the smallest real part ($|\lambda_3|> |{\rm Re}(\lambda_{1,2})|$) but that for $\rEP<\rH<\rBD$ the situation is reversed ($|\lambda_3|< |{\rm Re}(\lambda_{1,2})|$). Since generically the solution trajectory is expected to approach the origin along the eigenvector corresponding the eigenvalue(s) {\it closest} to the origin, we see that for $\rEP<\rH<\rBD$ the trajectory approaches the origin along a direction associated with a real eigenvalue, implying absence of spatial locking in this parameter regime. Thus, in the present system the transition from the presence of locking to its absence as $\rH$ increases occurs at $\rEP \simeq 1.04\cdot 10^{-5}$ instead of $\rBD \simeq 1.053\cdot 10^{-5}$ and we expect, therefore, that states consisting of identical peaks with unequal spacing can only exist in the interval $\rT<\rho_{\text{H}}<\rEP$. This result agrees with the continuation of the primary branches of $N$-peak states with large interpeak separations (see Fig.~\ref{fig:snaking}(b,c)), such as $N=1,2$ on $L=60$, for which the peak profiles develop oscillations once $\rH<\rEP\simeq  1.04\cdot 10^{-5}$ leading, in general, to locking and hence non-equispaced peak states (if $N>1$) as soon as $\rH<\rEP$. This is a significant new feature inherent in the present problem.
	
	We focus first on the bifurcations creating the $N=1$, 2, 3 and 4 branches near $\rH=\rT$, followed by the behavior near the right folds which are present in the region $\rH>\rEP$ before returning to a discussion of the behavior near the finite amplitude left folds which are also located in $\rT<\rH<\rEP$. 
	
	\subsection{Origin of the primary states}
	
\begin{figure}[!t] %F6
		\centering\includegraphics[width=0.75\textwidth]{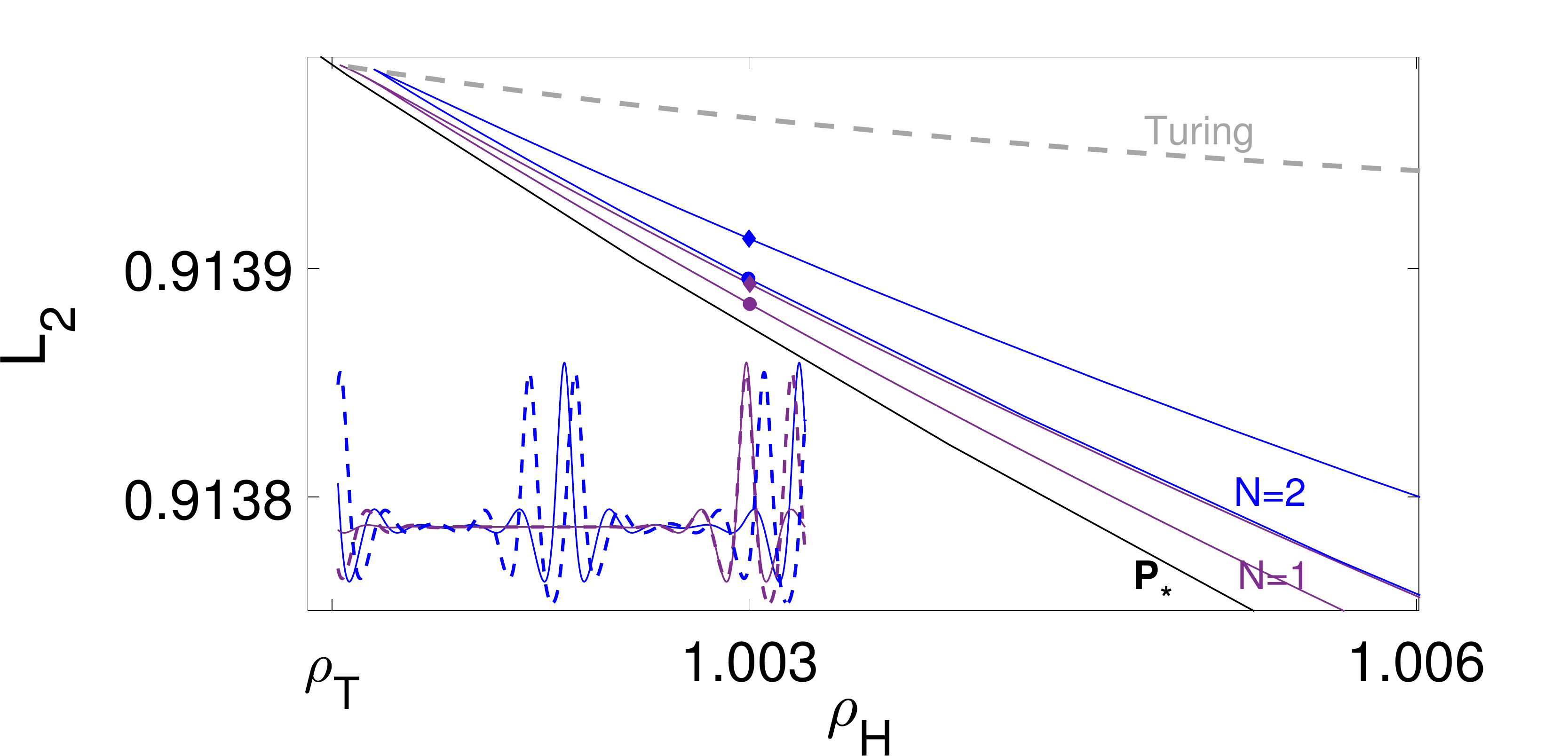}
		\caption{Origin of the $N$-peak states for $L=30$ showing the ${\text L}_2$ norm of the subcritical Turing branch together with a pair of secondary branches bifurcating from it at small amplitude, all as functions of $\rH$; $\vP_*$ denotes the homogeneous state. The first secondary states (purple curves) are the result of a modulational instability of the Turing state with period $L$ and hence are referred to as $N=1$ states. The lower of these develop into a single peak state, labeled $N=1$ in the figure, while the upper develops into an $N=1$ state consisting of a pair of adjacent peaks whose separation gradually increases with increasing $\rH$ until it reaches $L/2$ at $\rEP$ where it connects to an $N=2$ branch consisting of a pair of equispaced single peaks generated as a result of a modulational instability of the Turing state with period $L/2$ (lower blue curve labelled $N=2$). This instability also produces an $N=2$ state consisting of two groups of 2-peak states (upper blue curve) that terminates on an $N=4$ branch, also at $\rEP$. The inset shows the color-coded solution profiles on each branch at $\rH=1.003\cdot 10^{-5}$ (the solid profiles correspond to the lower branch of each pair, indicated by the $\bullet$ symbol, while the dashed profiles correspond to the upper branches, indicated by the $\blacklozenge$ symbol).} 
		\label{fig:origin}
	\end{figure}

	The dispersion relation in Fig.~\ref{fig:bif_uni}(b) indicates that on the real line the homogeneous state loses stability as $\rH$ decreases at a Turing bifurcation that takes place at $\rH=\rT$. This bifurcation creates a subcritical Turing pattern with wavelength $\ellc$~\cite{yochelis2020nonlinear}. In a periodic domain of finite length $L$ this bifurcation is slightly shifted as is the Turing wavelength $\ellc$ in order that the domain accommodate an integer number $n$ of wavelengths ($n=10$ when $L=30$). We refer to the resulting branch as the Turing branch. Because it is subcritical this Turing state undergoes a modulational instability already at very small amplitude generating a pair of modulated Turing states with modulation wavelength $L$. These two states differ in whether the maximum amplitude of the modulation coincides with a maximum of the Turing pattern or its minimum. Both of these branches generate states we refer to as $N=1$ states. Indeed, when followed from this modulation onset the former develops into a $N=1$ state consisting of a single peak while the latter develop into a $N=1$ state consisting of a pair of adjacent peaks, i.e., states consisting of a single peak and two peaks in the domain $L$ appear simultaneously via a secondary bifurcation from the Turing state, and so are strictly speaking secondary branches (purple curves in Fig.~\ref{fig:origin}). However, since this secondary bifurcation occurs at very small amplitude (and is invisible in large scale bifurcation diagrams, such as Fig.~\ref{fig:snaking}), we refer to these states in the following as primary states, as already done in Fig.~\ref{fig:snaking}.
	\begin{figure}[!t] %F7
		(a)\centering\includegraphics[width=0.75\textwidth]{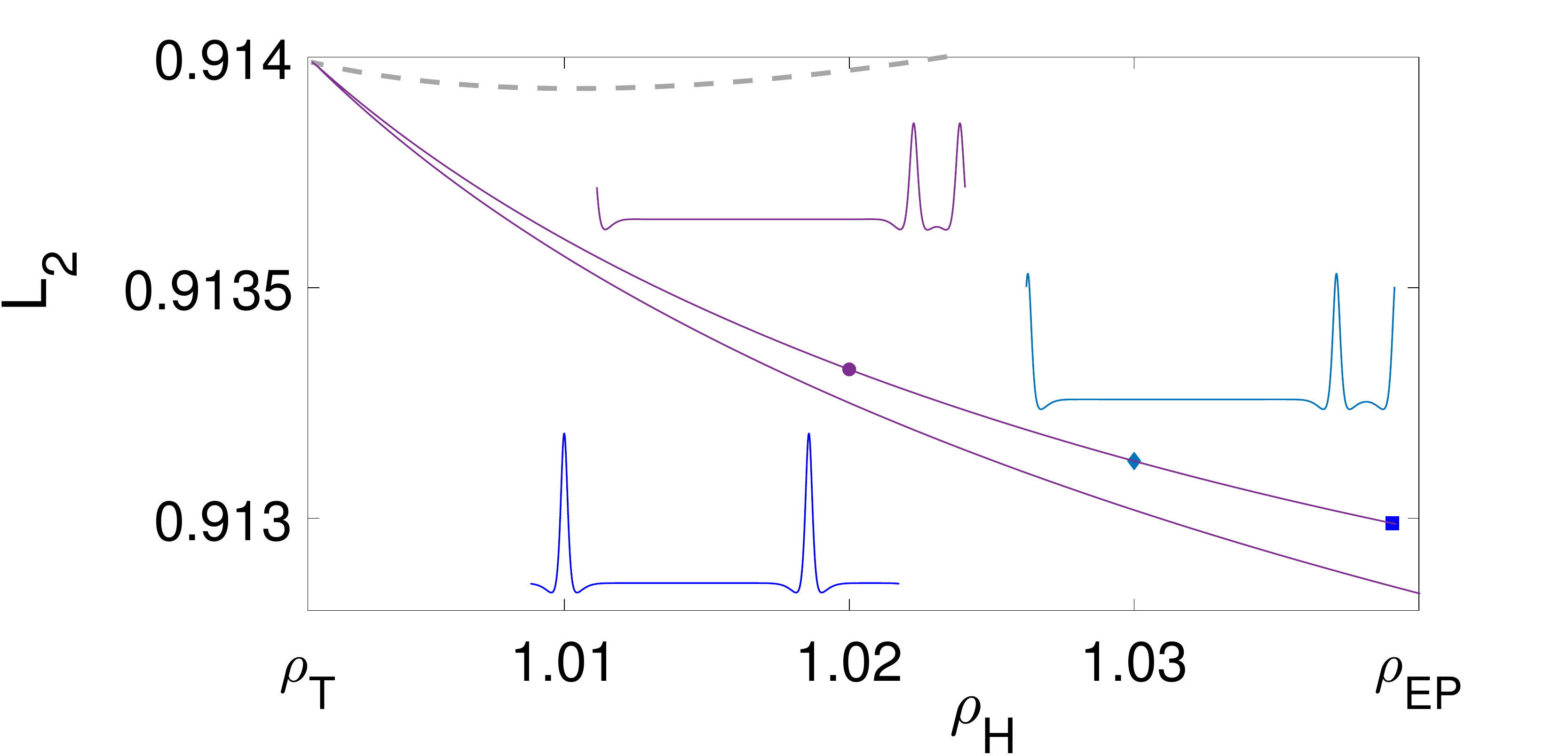}
		(b)\centering\includegraphics[width=0.75\textwidth]{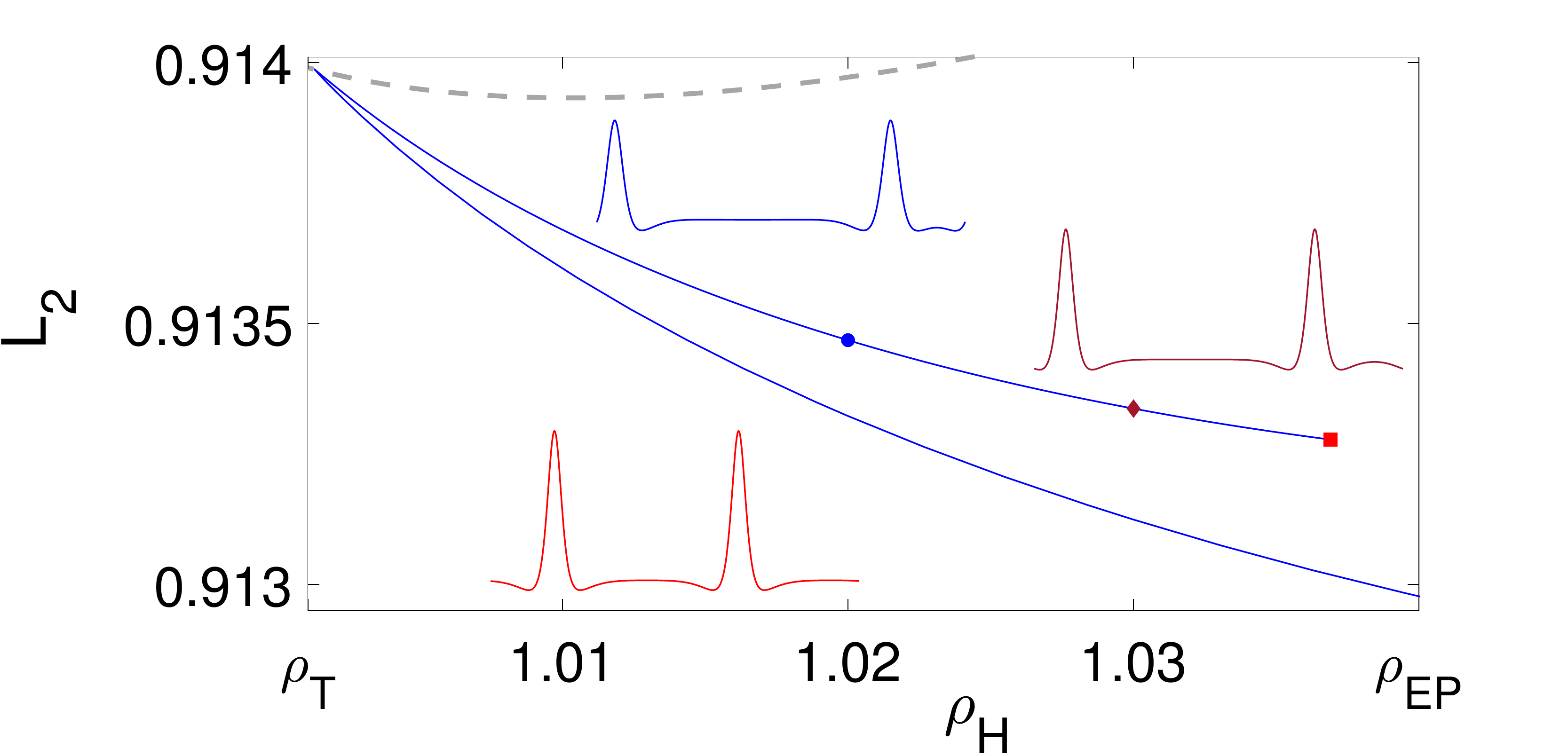}
		\caption{Color-coded profiles along the upper (a) $N=1$ and (b) $N=2$ branches in Fig.~\ref{fig:origin} farther from onset, showing the increasing distance between the peaks on these branches as $\rH\to\rEP$ and their separation approaches $L/2N$. In the $N=2$ case the profiles are shown on the half-domain $L/2$ only, $L=30$. The dashed line is the Turing branch.} 
		\label{fig:separation}
	\end{figure}
	
	The single peak $N=1$ state (lower purple branch) is shown in Fig.~\ref{fig:snaking}(a) over a much larger range of $\rH$. In contrast, the two-peak $N=1$ state (upper purple branch) has only a small extent: along the two-peak branch the separation of the two peaks gradually increases and approaches $L/2$ at $\rH\to\rEP$ (Fig.~\ref{fig:separation}). At this point the branch terminates on an $N=2$ branch consisting of a pair of identical equispaced states created in a subsequent modulational instability of the Turing state, this time with modulation period $L/2$ (lower blue curve in Fig.~\ref{fig:origin}). Because this period is shorter this instability takes place at a larger amplitude than the $N=1$ instability (see Fig.~\ref{fig:origin}). Like the lower $N=1$ branch this $N=2$ branch has a large extent in $\rH$ and is also shown in Fig.~\ref{fig:snaking}(a). Note that these modulational instabilities correspond to spatial $n:N$ resonances between the wavelength $L/n$ Turing pattern ($n=10$) and the modulation wavelength $L/N$, viz. $10:1$, $10:2$ etc. We shall see in the next section that spatial resonances also play an important role at the right folds of the surviving primary $N$-peak branches.
	
	We believe, but have not checked, that the other primary $N$-peak branches shown in Fig.~\ref{fig:snaking}(a) are generated via the same mechanism (modulational instability followed by spatial localization) as the $N=1,2$ branches discussed above; in each case the exchange point $\rH=\rEP$ serves to ``prune'' the branches emerging from successive modulational instabilities with wavelength $L/N$, $N<n$, leaving only equispaced states beyond $\rEP$. 
	
	We mention that true $N=1,2,\dots$ primary states are the result of primary bifurcations from the homogeneous state that occur as $\rH$ decreases {\it below} $\rT$, at parameter values where wavenumbers $k=2N\pi/L$ are themselves marginal ($\sigma=0$). For example, for our parameter values and $L=20$ the true primary $N=1$ state is created at $\rH\simeq 0.9903\cdot 10^{-5}$, a parameter value very close to the location of the left fold of the homogeneous state, $\rT\simeq 0.99\cdot 10^{-5}$. These states are completely distinct from the $N$-peak states created in the modulational mechanism just described and present in $\rH>\rT$. In particular, true primary states are present even when the Turing bifurcation is supercritical while the $N$-peak states studied here require the presence of a small amplitude modulational instability that in turn requires this bifurcation to be subcritical. Note that as the domain size increases all these secondary modulational instabilities collapse on $\rT$. Thus, when the Turing bifurcation is subcritical and the domain infinite, the Turing bifurcation is responsible for a large multiplicity of distinct states.

	\subsection{Behavior near the right folds}
	
	Each fold on the right (see Fig.~\ref{fig:snaking}(a)) represents a fold of a branch of identical equispaced peaks created as described in the preceding section. Near such folds there is always a Floquet multiplier that is close to $+1$, i.e., close to neutral stability with respect to amplitude perturbations. This fact allows, under appropriate circumstances, nearby bifurcations to secondary branches consisting of peaks with {\it nonidentical} heights. The number of new branches depends on the number $N$ characterizing the primary branch. Thus when $N=1$ there is no secondary branch of this type, while when $N=2$ there is one such branch consisting of one larger (L) peak and one smaller (S) peak, relative to the peaks in the primary solution. Note that in view of the periodic boundary conditions used an LS state is the same as an SL state. When $N=3$ there are two secondary branches, corresponding to LSS and LLS, while for $N=4$ there are 4 secondary branches, LSSS, LLSS, LSLS and LLLS. It is important to note that the resulting peaks are still present at locations very close to $x=(n/N)L$, $0\le n < N$, modulo overall translation, even though they are no longer identical. However, when the number $N$ of peaks within the domain becomes too large the behavior of the system starts to depart from the universal behavior exhibited in Fig.~\ref{fig:snaking}(a), see Fig.~\ref{fig:rfolds}.
	
	Figure \ref{fig:snaking}(a) shows details of this universal behavior. Each of the new states can be considered to be the result of spatial modulation of a periodic train of identical peaks associated with the onset of Eckhaus instability in the vicinity of the fold \cite{mercader}. The bifurcation that takes place closest to the fold corresponds to modulation with the longest possible wavelength, and so leads to states with one modulation wavelength in the domain; subsequent modulational instabilities generate states exhibiting two modulation wavelengths in the domain such as the state LSLS but these occur farther away from the fold. This is essentially the same as the mechanism responsible for the presence of the $N$-peak states in the first place, except that here the relevant spatial resonances are of lower order since $N$ is much smaller than $n$, the number of Turing wavelengths that fit in the domain. However, it is still true that the long wave states are generated in $N:1$ spatial resonances that take place nearest to the fold, while the other states mentioned are the consequence of additional spatial resonances that are present farther from it \cite{mercader}.
	
	We mention that the local behavior near a $N:1$ spatial resonance is described by amplitude equations of the form \cite{dangelmayr}
	\begin{equation}\label{eq:resonance}
		{\dot B}_N=\mu B_N + c_1{\bar B}_N^{N-1}+c_2|B_N|^2B_N,    
	\end{equation}
	where $B_N$ is a complex modulation amplitude such that $|B_N|$ measures the height difference between L and S, $\mu$ represents the distance from the resonance and $c_1$, $c_2$ is a pair of real constants; ${\bar B}_N$ is the complex conjugate of $B_N$. Writing $B_N=r_N\exp i\phi_N$ we see that the new branches generated in the $N:1$ resonance correspond to solutions of $\sin N\phi_N=0$, or $\phi_N=\pi m/N$, $m=0,1,\dots,N-1$. Not all of these will have distinct ${\text L}_2$ norms, however. Note also that the strong resonances $N=2$ and $N=3$ differ from the rest. In particular, when $N=2$ there is only one new state, LS say, corresponding to $m=1$, while the $N=3$ case leads to a transcritical bifurcation to LSS and LLS, corresponding to $m=1$ and $m=2$, as seen in Fig.~\ref{fig:snaking}(a). Note that this is (locally) a transcritical bifurcation only because the bifurcation does not in fact coincide with the fold, although it is very close to it.
	\begin{figure}[!t] %F8
		(a)\centering\includegraphics[width=0.75\textwidth]{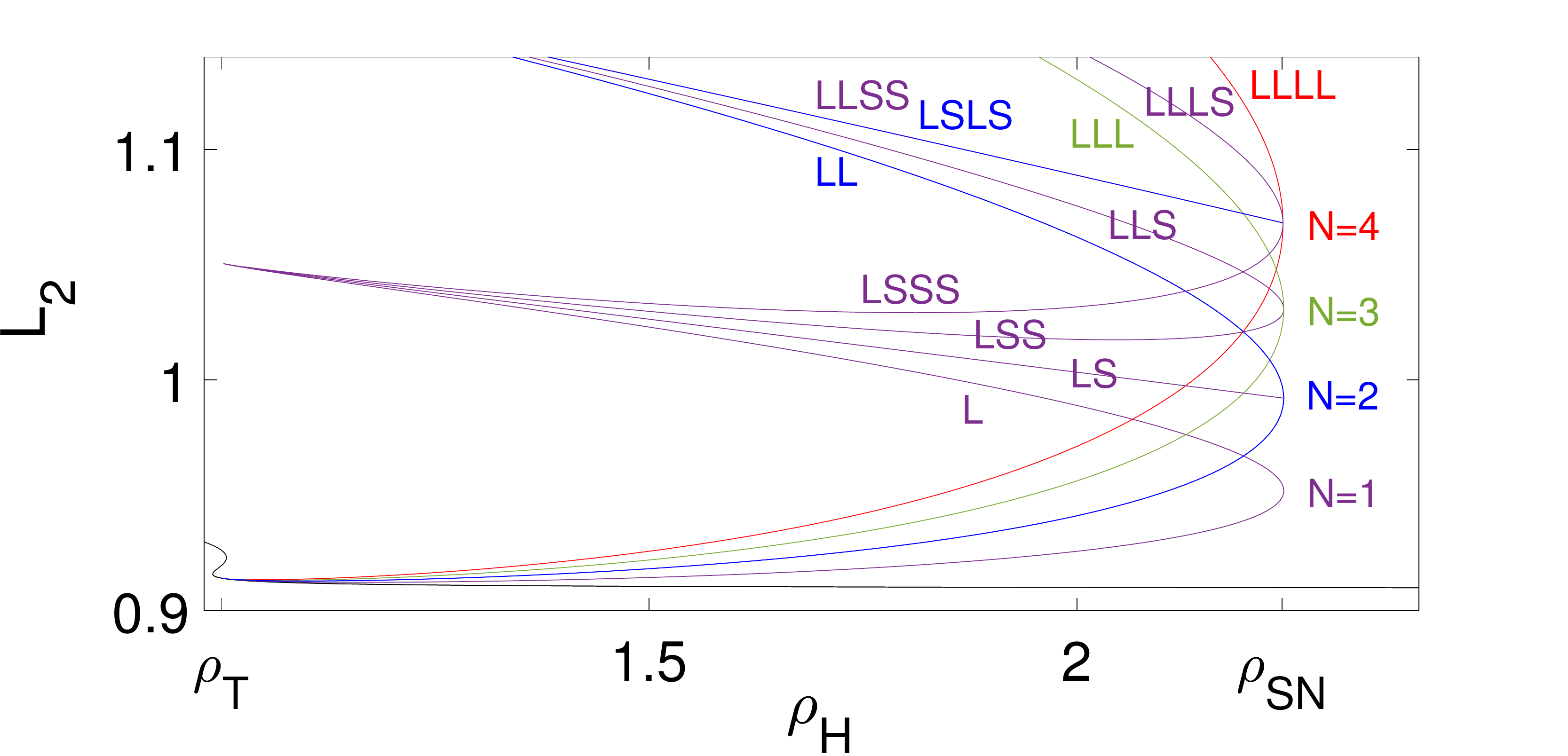}
		(b)\centering\includegraphics[width=0.7\textwidth]{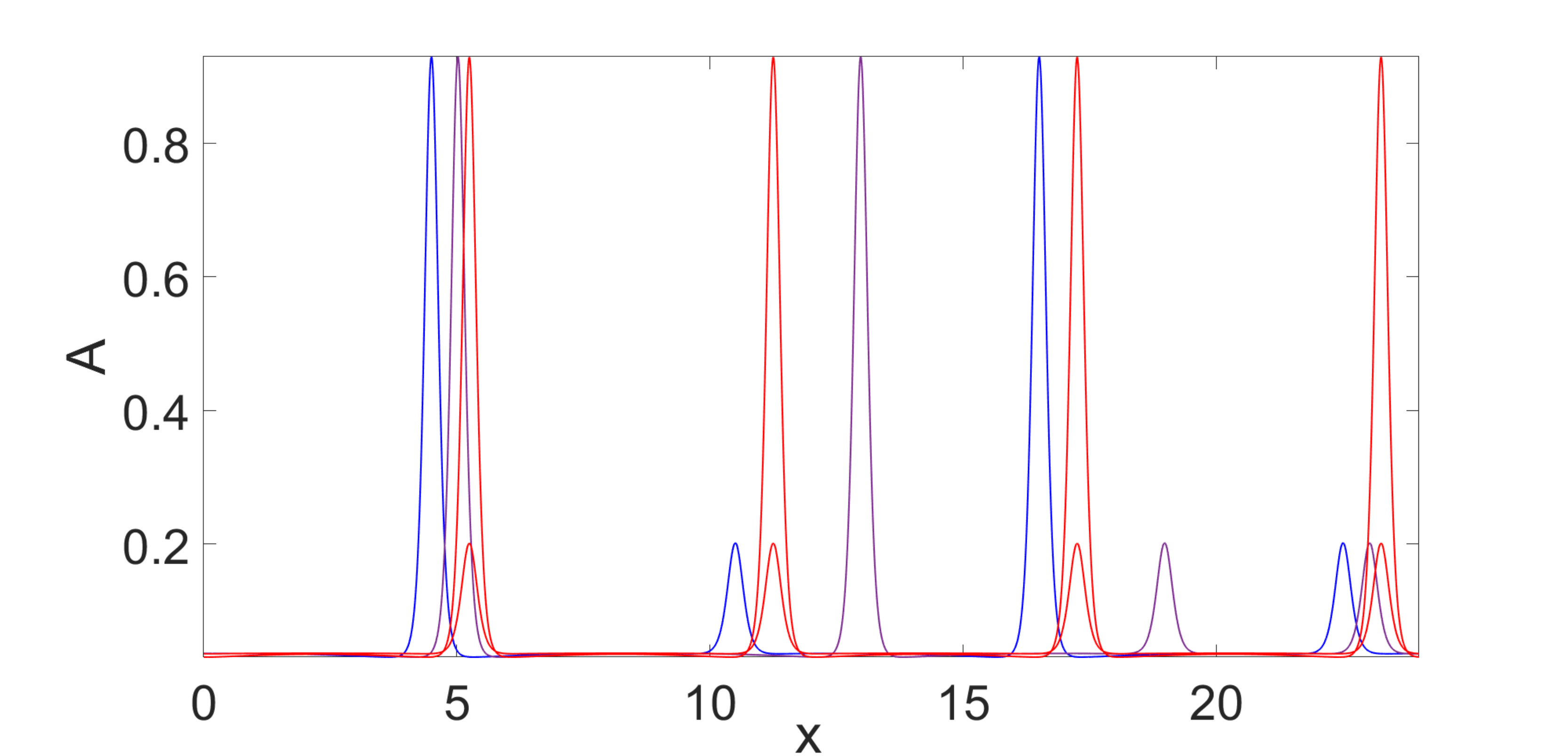}
		(c)\centering\includegraphics[width=0.7\textwidth]{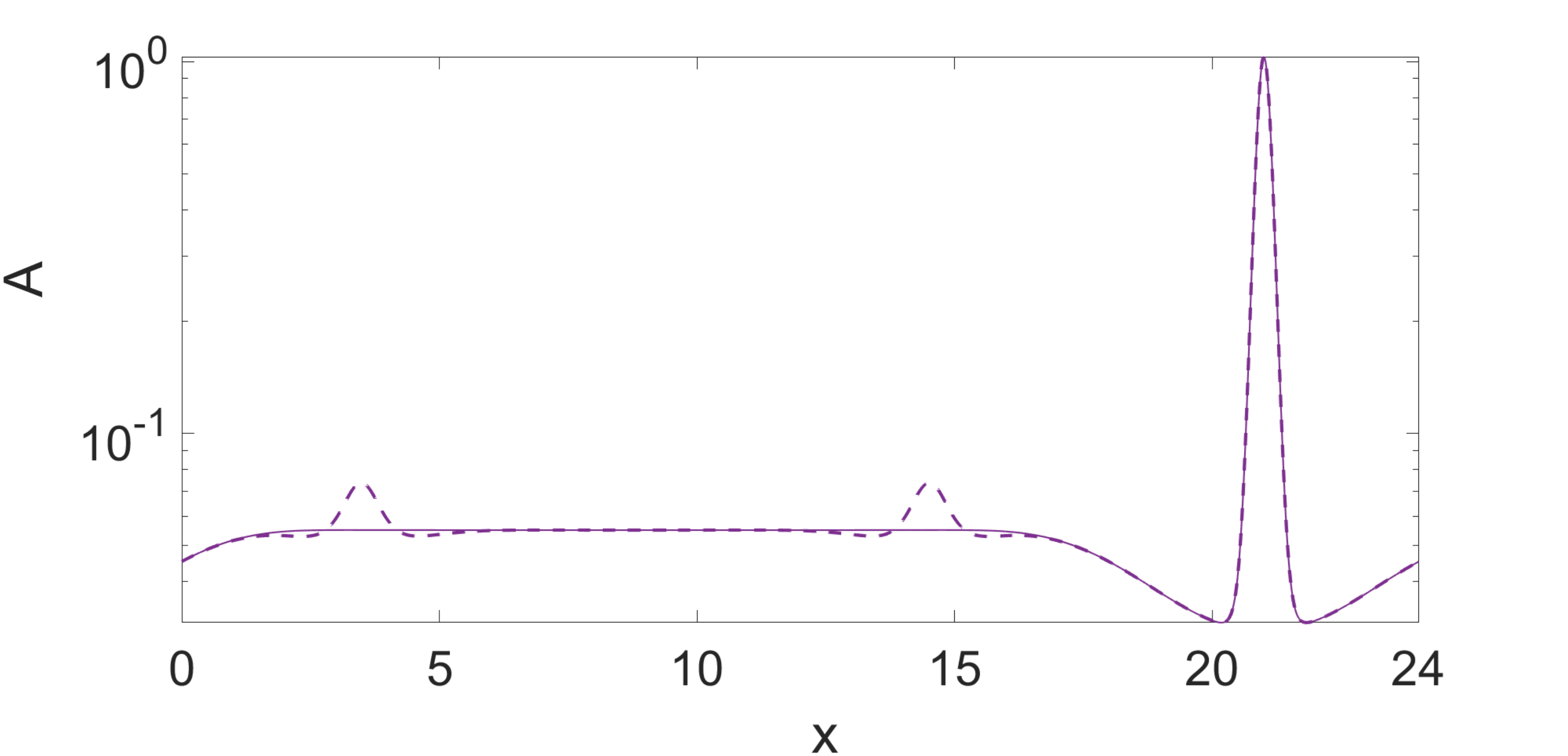}
		\caption{(a) Details of the bifurcation diagram for $L=24$ near the folds on the right confirming the universality of the behavior shown in Fig.~\ref{fig:snaking}(a). (b) Solution profiles along the $N=4$ branch at $\rH=1.5 \cdot 10^{-5}$. Blue: LSLS. Purple: LLSS. Red: coexisting LLLL and SSSS states on the $N=4$ branch of identical equispaced peaks. Relative position of the peaks is a property of each solution; in the LLSS state the SS peaks are significantly closer to one another than the LL peaks. (c) Profiles along the branch connecting L to LSS via the upper left fold, both at $\rH=1.02\cdot 10^{-5}$. L: solid line; LSS: dashed line. The profiles are shown on a logarithmic scale to evince the pair of small peaks S present in the latter case.}
		\label{fig:profs4}
	\end{figure}
	
	The case $N=4$ is also special. Here all the new branches bifurcate in the same direction relative to the bifurcation point $\mu=0$. The above theory predicts two distinct states with modulation wavelength $L$, but there are in addition branches exhibiting modulation on smaller scales that also bifurcate very close to the fold although slightly farther from it. Given that there are four peaks in the domain, the possible modulated states are LLSS, LSLS, LLLS and LSSS. All of these have been found and are included in Fig.~\ref{fig:profs4}(a), cf.~\cite{Lojacono2017}. Note that since the ${\text L}_2$ norm of the LLSS and LSLS states is essentially the same the diagram appears to show 3 distinct branches bifurcating from the $N=4$ fold (more generally we expect $N-1$ such distinct branches). The LLLS and LSSS states are expected for any domain length but are omitted from Fig.~\ref{fig:snaking}(a) where $L=60$. 
	
	We have checked (Fig.~\ref{fig:profs4}(b)) that at each $\rH$ to the left of this fold the peak profiles present in the observed unequal peak states do indeed accurately match the peak profiles on the upper and lower branches of the $N$-peak states present at this value of $\rH$. Observe that in the LLSS state the SS peaks are significantly closer to one another than the LL peaks. We can understand this property in terms of a force between adjacent peaks when $\rH>\rEP$. In this regime, the preferred state is the $N=1$ in domains of any length. If a second identical peak is inserted the peaks space themselves as far as possible from each other, i.e., one obtains an $N=2$ primary state. Evidently the peaks {\it repel} one another~\cite{yochelis2008formation,verschueren2017}, and so inserting further peaks leads to $N=3,4,...$ primary states, all consisting of identical equispaced peaks. The repulsive force depends on the amplitude of the peaks, however, and is stronger between taller peaks and weaker between shorter peaks. This fact accounts for the different separations within the LLSS state and the LSLS state; these separations depend on $\rH$ which in turn determines the height of the L and S peaks and hence the force between adjacent LL, LS and SS peaks. This is so for LLLS and LSSS as well (not shown here, but see Fig.~\ref{fig:unequalprofs60}(c) for $L=60$). The interpeak potential can be computed as for the Ginzburg-Landau equation \cite{kawasaki1982} but we do not do this here.
	
	Based on the above results we expect $N$-peak states of the form SL$...$L and LS$...$S to be present for any $N$. When $N$ and the domain length $L$ are both large, these states approximate homoclinic connections from L$...$L to itself and from S$...$S to itself, respectively, while states such as S$...$SL$...$L represent heteroclinic cycles connecting S$...$S to L$...$L and back again, that is, connections between the small and large $N$-peak states that coexist at each $\rH$ to the left of the right fold of the $N$-peak state. Thus we expect periodic trains of peaks to exhibit the same localization and subsequent snaking as that arising from the subcritical Turing bifurcation itself. 
	
	Note that from the point of view of the primary $N=2$ state the bifurcation generating the state LS is a spatial period-doubling bifurcation, while that generating the LSS and LLS states from $N=3$ is a spatial period-tripling bifurcation, etc. There are in general two distinct ways to generate a spatial period-doubling bifurcation: such a bifurcation can be due to the appearance of a difference in separation between adjacent peaks (as occurs near $\rEP$ for the small amplitude terminating branches) or due to a difference in the heights of adjacent peaks, as occurs at the right folds $\rSN$). A similar distinction between phase and amplitude-generated period multiplication applies for other $N:1$ resonances.

	\subsection{Behavior near the upper left folds}
	
\begin{figure}[!t] %F9
		(a)\centering\includegraphics[width=0.75\textwidth]{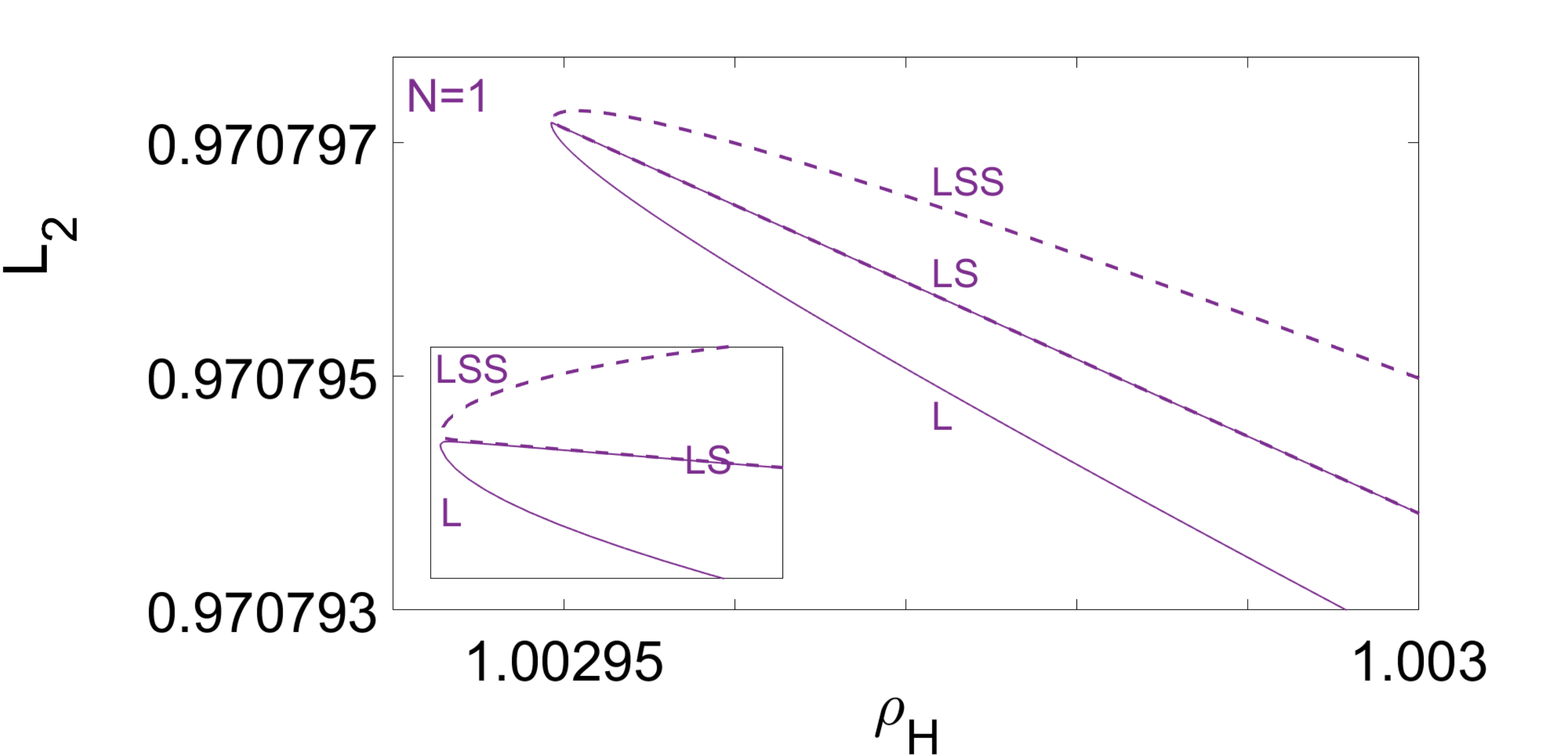}
		(b)\centering\includegraphics[width=0.75\textwidth]{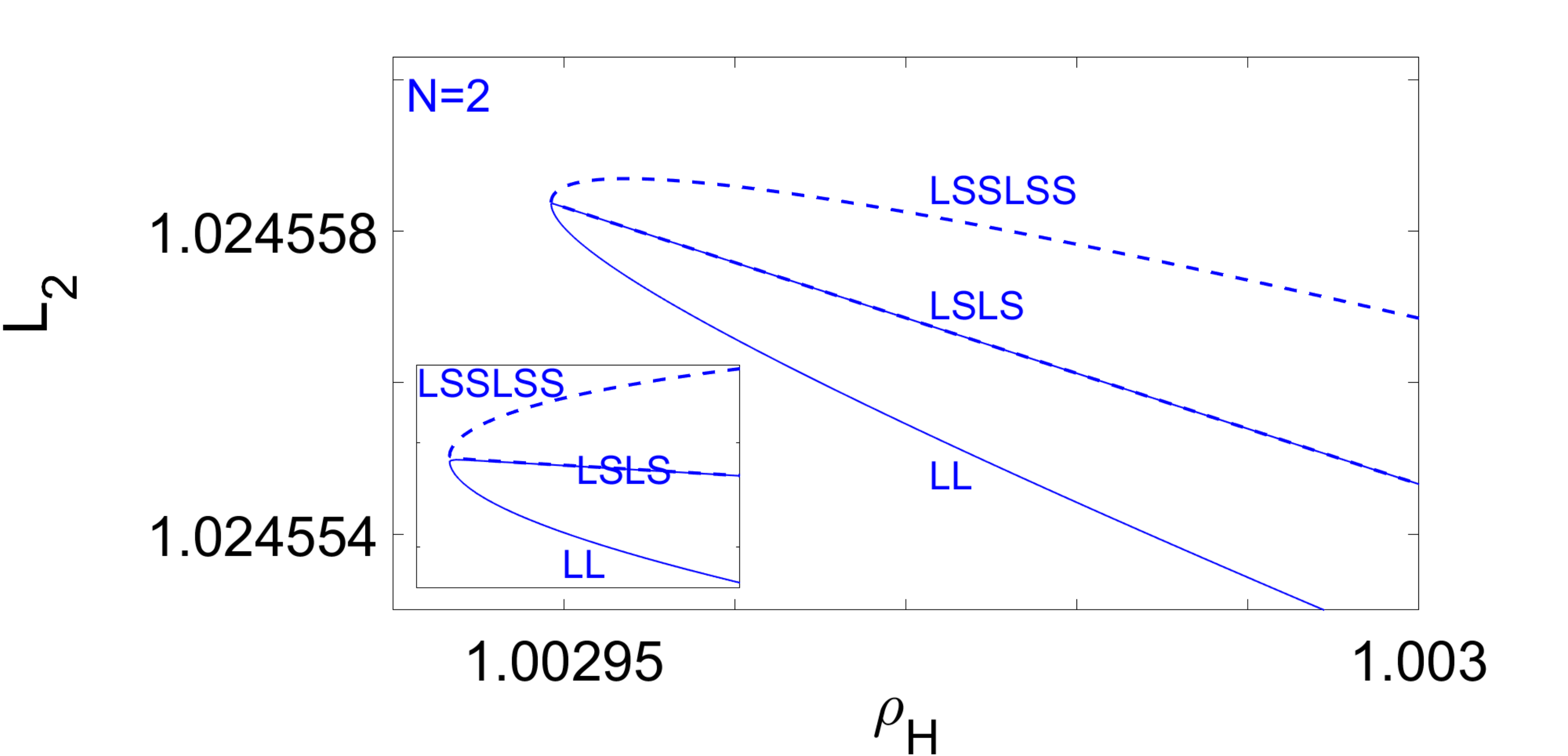}
		\caption{Detail of the $L=60$ bifurcation diagram near the upper left for (a) $N=1$ and (b) $N=2$, showing that the cusp-like features in Fig.~\ref{fig:snaking} are in fact narrow folds with secondary bifurcations generating LS and LSLS branches. These folds are very close to $\rT\simeq 1.0011 \cdot 10^{-5}$, the Turing bifucation that creates the $N$-peak states at the lower left of the full bifurcation diagram.}
		\label{fig:cusp}
	\end{figure}

	As already mentioned, the apparent termination points of all multipeak branches on the left (Fig.~\ref{fig:snaking}(b,c) for the case $L=60$) are in fact sharp, cusp-like folds (see Fig.~\ref{fig:cusp}(a,b)). Similar behavior is present for $L=24$ (Fig.~\ref{fig:profs4}(a)) as well as other domain sizes. We believe that these folds accumulate on $\rH=\rT$, with $N=1$ closest, $N=2$ a little farther away etc., a consequence of the fact that the leading spatial eigenvalues of the homogeneous state are purely imaginary for $\rH<\rT$. Near these folds each primary branch transitions into a state of nonidentical peaks. At the right folds this occurs when the periodic state starts to develop differences in the peak heights, a transition that sets in via bifurcations generated by spatial resonances. On the upper left the transition is instead continuous and occurs when new peaks start to grow in the interpeak region between existing peaks. Figure \ref{fig:profs4}(c) shows the transition from a 1-peak state L to a 3-peak state LSS as one passes the left fold when $L=24$: two identical new small peaks appear in the domain below the fold and grow in height as one passes the fold and beyond. Owing to the periodicity of the domain these small peaks may be thought of a growing sidebands of the main peak, and hence as an SLS state \cite{verschueren2018dissecting}.
	
	Near the (upper) left fold the peaks in the resulting LSS (SLS) state are not equispaced, but as one follows the LSS branch through $\rEP$ and beyond, the amplitude of the small peaks continues to grow and their location changes in response. By the end of the branch the three peaks are identical and equispaced, and the branch terminates in a 3:1 resonance near the right fold on the 3-peak primary branch that emanates from the vicinity of the Turing bifurcation. The figure also shows a third branch, labeled LS. This branch appears to bifurcate from the upper left fold on the L branch and is characterized by the appearance of only a single new peak (Fig.~\ref{fig:cusp}(a)). In fact, as shown in the inset in Fig.~\ref{fig:cusp}(b) the LS branch is split into two almost identical branches, one consisting of L and a small peak S a distance $L/2$ from it that connects smoothly with the L branch and a second LS branch with S at approximately $L/3$ from L. This branch connects smoothly to LSS. The exponentially small splitting between these two branches is the result of slightly different interactions between the L and S peaks in these two nominally identical states. Like the LSS branch, the former LS branch can be continued all the way to its termination near the right fold of the primary $N=2$ branch -- at the termination both peaks are of the same height and equispaced -- but this is not the case for the second LS branch which must terminate by $\rEP$. Similar behavior occurs on the upper left for the $N=2$ states (Fig.~\ref{fig:cusp}(b)).
	
\begin{figure}[!t] %F10
		(a)\centering\includegraphics[width=0.75\textwidth]{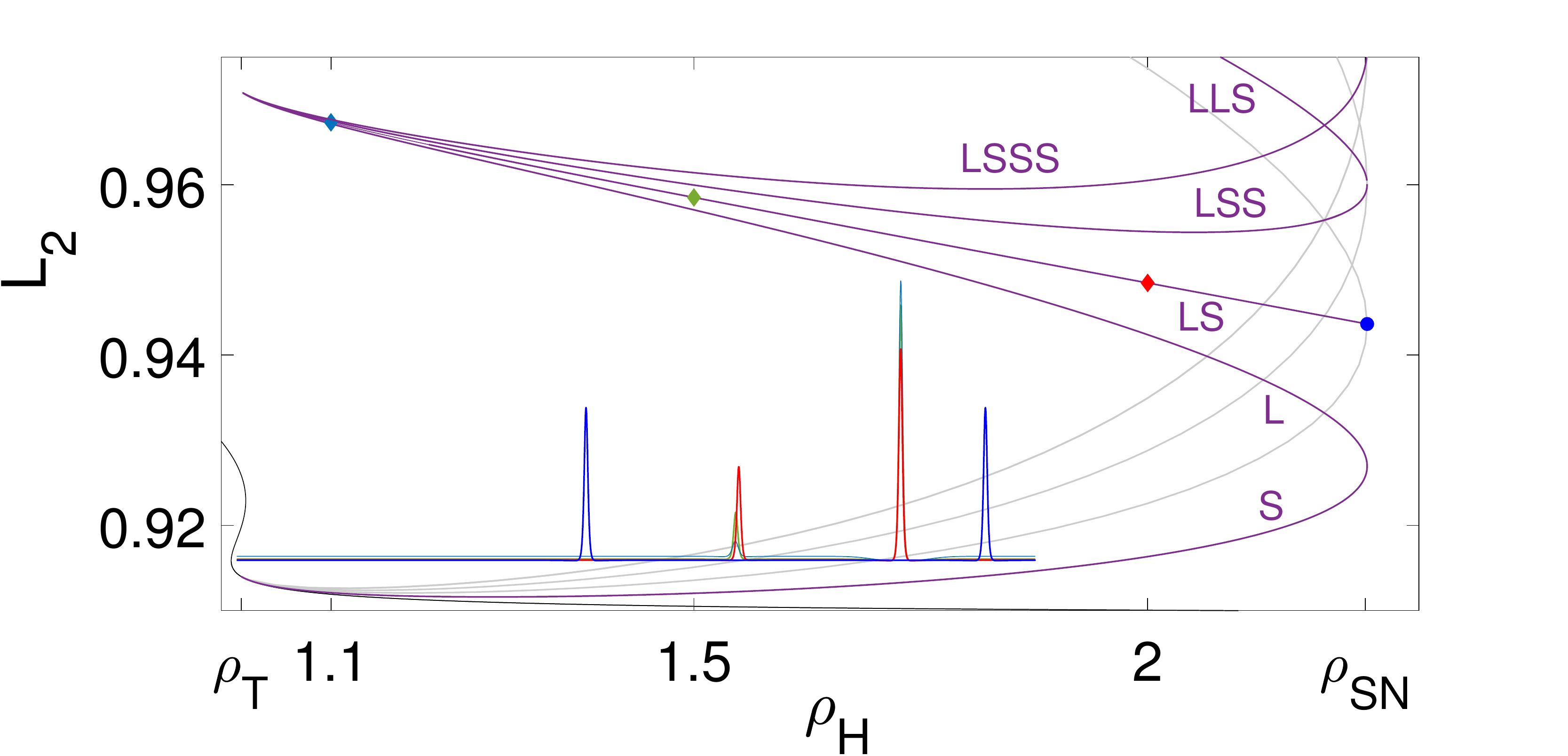}
		(b)\centering\includegraphics[width=0.75\textwidth]{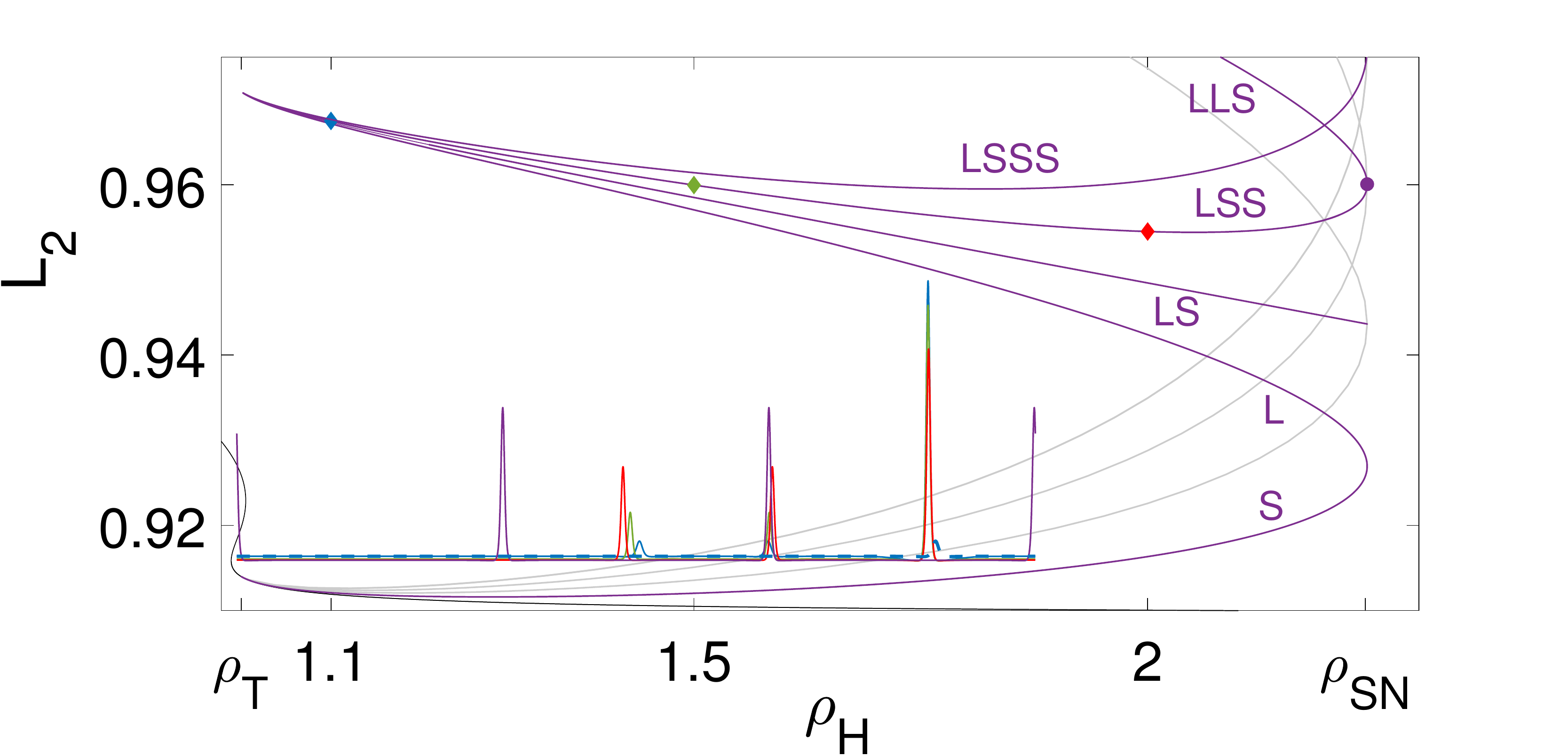}
		(c)\centering\includegraphics[width=0.75\textwidth]{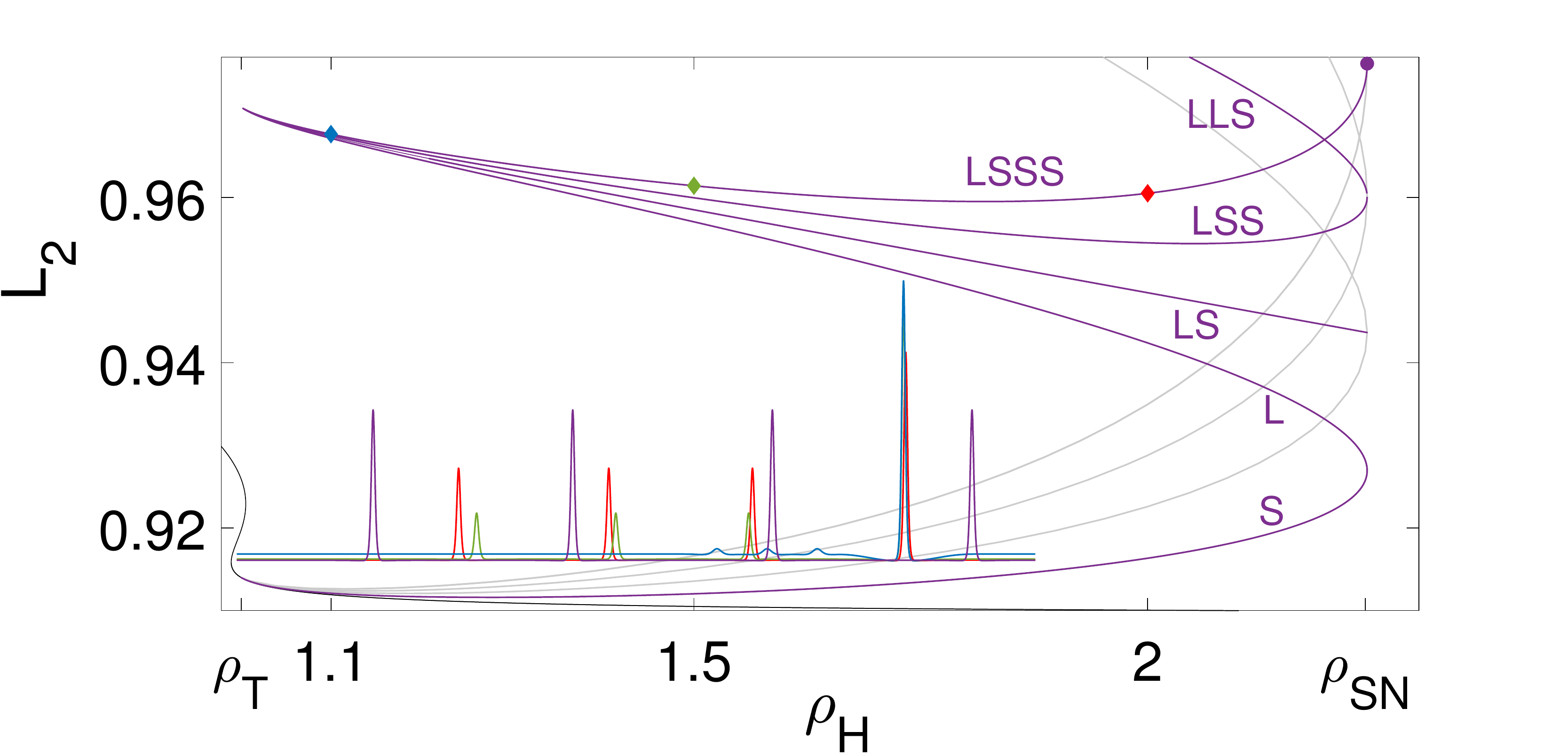}
		\caption{Evolution of the peak heights and spacing along (a) the SL branch, (b) the SSL branch and (c) the SSSL branch created in the upper left $N=1$ fold when $L=60$. These branches terminate on the primary 2-peak, 3-peak and 4-peak branches, respectively. The peaks separate rapidly near their termination in order to reach the respective separations $L/2$, $L/3$ and $L/4$.}
		\label{fig:unequalprofs60}
	\end{figure}

	We mention that for larger values of the domain length $L$ the numerical continuation of the LSS branch (and related branches) through the exchange point $\rEP$ as $\rH$ increases becomes exceedingly difficult and one is tempted to conclude that the branch terminates at $\rEP$. While such a termination is indeed possible on the real line where the peaks can move arbitrarily far apart as one approaches $\rEP$ this is not the case on finite domains. Instead, we find that while the peaks do indeed start to move apart as $\rH$ increases the solution branch may pass smoothly through $\rEP$ and extend all the way to the right fold of the $N=2,3,\dots$-peak states where all $N$ peaks are of the same height and equidistant (Fig.~\ref{fig:unequalprofs60}), provided only that the different peaks are already approximately equispaced below $\rEP$. In fact, we expect additional branches in $\rH<\rEP$ with S peaks separated from L by different numbers of Turing wavelengths that do not extend beyond $\rEP$ and the number of such branches will increase rapidly with the domain length $L$. Thus here, too, the EP point serves to prune solution branches. We believe this is the reason why numerical continuation with $\rH$ {\it decreasing} through $\rEP$ remains possible even when continuation in the opposite direction becomes impossible, an important fact that appears to have been missed in earlier work~\cite{verschueren2017}.
	
	The above discussion explains why the interval $\rT<\rH<\rEP$ is so special: it permits spatial locking owing to the presence of oscillations in the tails of the peaks while for $\rH>\rEP$ the peak profiles become monotonic and spatial locking is absent. Our numerical continuation suggests that the new peaks that appear along the $N=1,2,\dots$ branches near the upper left folds arise from the growth of select peaks in these tails, explaining why the $N=1,2,\dots$ states start adding new peaks only once they enter $\rH<\rEP$. Figure~\ref{fig:profs4}(c) shows the creation of the LSS branch near the upper left fold when $L=24$. When the branch recrosses $\rEP$ the original peak and the new peaks unlock, but do not become equispaced. As already explained, this is a consequence of unequal forces between small and large peaks. Once all the peaks have acquired the same height, the peaks become equispaced and connect to the corresponding primary $N$-peak branch (Fig.~\ref{fig:unequalprofs60}), provided, of course, the domain length is sufficiently large ($L\gg \ellc$).

	\subsection{Foliated snaking}
	
	The behavior described in the two preceding subsections resembles a phenomenon called foliated snaking that was originally described in~\cite{ponedel2016forced} in a spatially forced spatially two-dimensional system, and subsequently identified in the homogeneously forced Lugiato-Lefever equation in Ref.~\cite{parra2018bifurcation}; see also \cite{glasner2012characterising}. However, the present system differs in two important ways. In our system the foliated snaking structure is created from successive modulational instabilities of a subcritical Turing state. In contrast, the $N$-peak states present in the Lugiato-Lefever equation are created at a fold of the uniform states \cite{parra2018bifurcation} while similar states present in both vegetation~\cite{ruiz} and Gierer-Meinhardt~\cite{yochelis2008formation} models are the result of a transcritical bifurcation of this state. Thus, our system is closer to the systems studied in~\cite{lloyd2013,verschueren2017,verschueren2018dissecting} rather than to the Lugiato-Lefever equation. Moreover, it is higher-dimensional than these examples and this fact is reflected in the behavior we observe.
	
	Foliated snaking is found in the present problem because the localized states created in the modulational instability of the Turing state extend into the region of no spatial locking where the leading spatial eigenvalues $\lambda$ are real -- in contrast to the Swift-Hohenberg equation in which the localized states remain confined to the locking region where the eigenvalues $\lambda$ form a quartet in the complex plane. We have seen that outside this region the localized states take the form of $N$-peak states, i.e. a periodic trains of equispaced stationary peaks characterized by the number $N$ of peaks in the domain.
	
	In general, we expect an absence of locking between adjacent peaks when $\rH>\rBD$, i.e., beyond the so-called Belyakov-Devaney transition, where the four leading spatial eigenvalues $\lambda$ are real. As a result we expect only equispaced peaks when $\rH>\rBD$ and non-equispaced peaks for $\rH<\rBD$. In other words we expect a large variety of states when $\rH<\rBD$ owing to the presence of spatial locking but a much simpler situation for $\rH>\rBD$ where locking is absent. This is the picture that is consistent with the behavior observed in 4-dimensional systems~\cite{parra2018bifurcation,verschueren2018dissecting}, where the role played by the BD point is reliably established.
	
	The present system is 6-dimensional, however, if we ignore the pair of large real eigenvalues. Figure \ref{fig:eigs} shows that for $\rT<\rH<\rEP<\rho_{BD}$ the complex eigenvalues have the smallest real part ($|\lambda_3|> |{\rm Re}(\lambda_{1,2})|$) but that for $\rEP<\rH<\rBD$ the situation is reversed ($|\lambda_3|< |{\rm Re}(\lambda_{1,2})|$). Since generically the solution trajectory approaches the origin along the eigenvector corresponding to the eigenvalue(s) {\it closest} to the origin, we see that for $\rEP<\rH<\rBD$ the trajectory approaches the origin along a direction associated with a real eigenvalue, implying absence of spatial locking in this parameter regime. Thus, in the present system the transition from the presence of locking to its absence as $\rH$ increases occurs at $\rEP\simeq 1.04\cdot 10^{-5}$ instead of $\rBD\simeq 1.053\cdot 10^{-5}$ and we, therefore, expect states with unequal peak spacing to exist in the interval $\rT<\rH<\rEP$ only. Our results are fully consistent with these conclusions.
	
	We now address a key issue, the relation between existing theory describing the unfolding of a global bifurcation at the BD point \cite{verschueren2018dissecting} and computations that are necessarily performed on finite (periodic) domains. On the real line identical peaks are expected to separate indefinitely as one approaches $\rEP$ from below, corresponding to the formation of an infinite number of homoclinic orbit(s) to the homogeneous state \cite{champneys1998homoclinic}. Thus, all primary $N$-peak ($N>1$) states would be expected to terminate at $\rEP$. In a finite periodic domain, however, a global bifurcation cannot take place and all states remain periodic. As a result some of the solutions created through the appearance of S peaks near the upper left fold can be followed through $\rEP$ and these become equal amplitude equispaced states by $\rSN$, since these states have (approximate) period $L/N$ we see that the passage through $\rEP$ in the direction of increasing $\rH$ changes the solution period from $L$ to $L/N$, i.e., the passage through this point is associated with wavenumber multiplication. Moreover, multipeak states exist in the whole interval $\rT<\rH\le\rSN$ instead of the expected interval $\rEP<\rH\le\rSN$ (the single peak state $N=1$ is expected to extend to $\rT$) and the BD point is no longer associated with global bifurcations. Other states, composed of dissimilar peaks with different separations necessarily terminate by $\rEP$ and cannot be continued past this point. These results are computationally difficult to establish unless one follows the branches of equispaced states from their creation at the right folds down in $\rH$, in other words, in the reverse direction. We conjecture that similar behavior is present in four-dimensional systems but may have been missed for this reason.
	
	We mention, finally, that branches such as the LS branch appear to be absent from the existing unfolding of the global bifurcation at the BD point~\cite{verschueren2018dissecting}. We have seen that such states in fact play a fundamental role in the overall structure of the foliated snaking bifurcation diagram identified here.
	
	\subsection{Composite-peak isolas}
	
	Although the structure we have uncovered so far is already quite complex, it omits a large variety of interesting additional states. From studies of the quadratic-cubic Swift-Hohenberg equation it is known that the snaking or pinning region is populated by a variety of so-called multipulse states \cite{Burke2009}. The simplest of these are equispaced 2-pulse states, consisting of two copies of a $1, 2, 3,\dots$-peak state, separated by $L/2$. States of this type lie on a connected branch that snakes much like the 1-pulse states, except that the number of back-and-forth oscillations is half that of the 1-pulse branch. However, the two pulses need not be separated by $L/2$: they can form bound groups, in which the pulses are separated by an integer number $p$ of half-wavelengths $\ell$. For each $p$ a 2-pulse state consisting of two identical pulses, each with $q$ peaks, lies on a closed branch called an isola, one for each pair of integers $(q,p)$. The smallest isola corresponds to $p=1$ and the isolas grow in size as $p$ increases, accumulating on the snakes-and-ladders structure of the 1-pulse bifurcation diagram in the pinning region, with higher $q$ isolas corresponding to higher $\text{L}_2$ norm. Thus, the snaking or pinning region contains a stack of nested isolas, one for each choice of $(p,q)$. This is not all, however, since bound states of two different pulses are also possible. These also snake (see, e.g., Fig.~6, of~\cite{Burke2009} where the snaking of a bound state consisting of one 1-peak state and one 2-peak state is investigated). Similar 3-pulse states etc are also present, altogether generating a structure of remarkable complexity.
	
	With this in mind we recall that for $\rT<\rH<\rEP$ the homogeneous state $\vP_*$ of the present system has leading eigenvalues $\lambda$ that form a quartet in the complex plane, and in this region we expect the present system to behave much like the Swift-Hohenberg equation. In particular, we expect that in addition to the $N$-pulse equispaced states discussed thus far, the system also exhibits multipulse states consisting of groups of pulses with different intergroup separations, and that these are created in the interval $\rT<\rH<\rEP$ in the same way as the corresponding states of the Swift-Hohenberg equation. Of course, each of these states will exit the interval $\rT<\rH<\rEP$ and so participate in the foliated snaking structure of the primary 1-pulse states. In the following we present the results of detailed computations of one such sequence of states, fully cognizant of the limitations of numerical studies of this kind in revealing the full complexity of the system. We think of the branches of multipulse states we find as providing decoration of the basic foliated structure already described.
	\begin{figure}[!t] %F11
		\centering\includegraphics[width=0.75\textwidth]{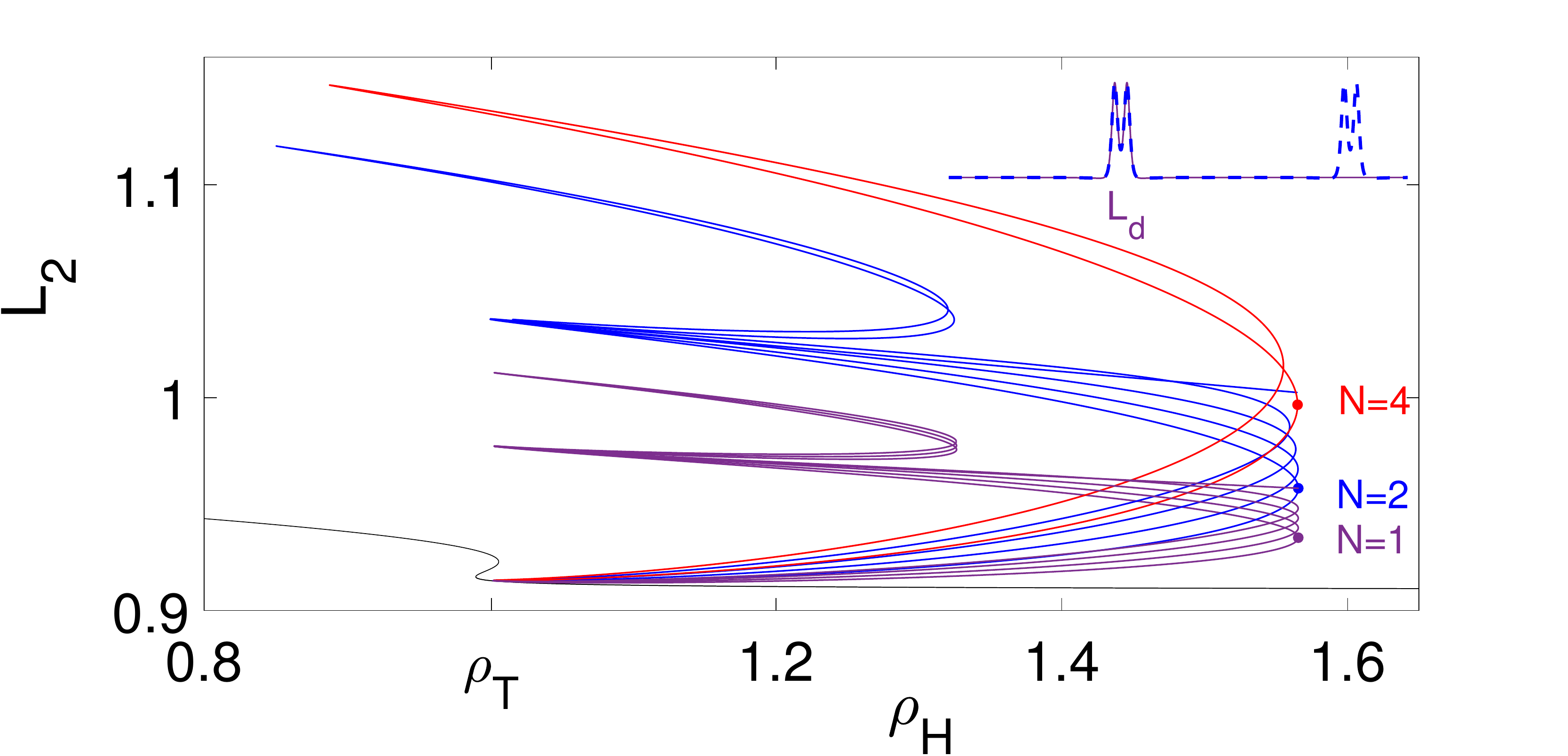}
		\caption{Bifurcation diagram for a subset of the multipulse states present in the system when $L=24$. Here the integers $N=1,2,\dots$ refer to branches of $N$-pulse states consisting of $N$ 2-peak groups. Examples of the $N=1,2$ states at the respective folds (dots) are included in the inset. The foliated snaking structure discussed in earlier figures (e.g., Fig.~\ref{fig:profs4}(a)) has been removed for clarity. Details of this diagram are described in the following two figures.}
		\label{fig:overall_split}
	\end{figure}
	
	Figure \ref{fig:overall_split} provides a summary of the states computed on $L=24$ and consisting of groups of 2-peak and 3-peak states. We first mention the branches labeled by $N=1,2,\dots \,$. In contrast to the primary branches of equispaced 1-peak states forming the skeleton of the foliated snaking structure (not shown in this figure), these branches consist of $N$ equispaced groups of two pulses. Profiles of $N=1$ and $N=2$ states of this type on $x\in [0,L]$ are shown in the inset using solid and dashed lines, respectively.
	
	The basic structure of the figure recapitulates that of Fig.~\ref{fig:profs4}(a), albeit with more branches. We now discuss the behavior shown in this diagram in greater detail, focusing again on the folds.
	
\begin{figure}[!t] %F12
		(a)\centering\includegraphics[width=0.75\textwidth]{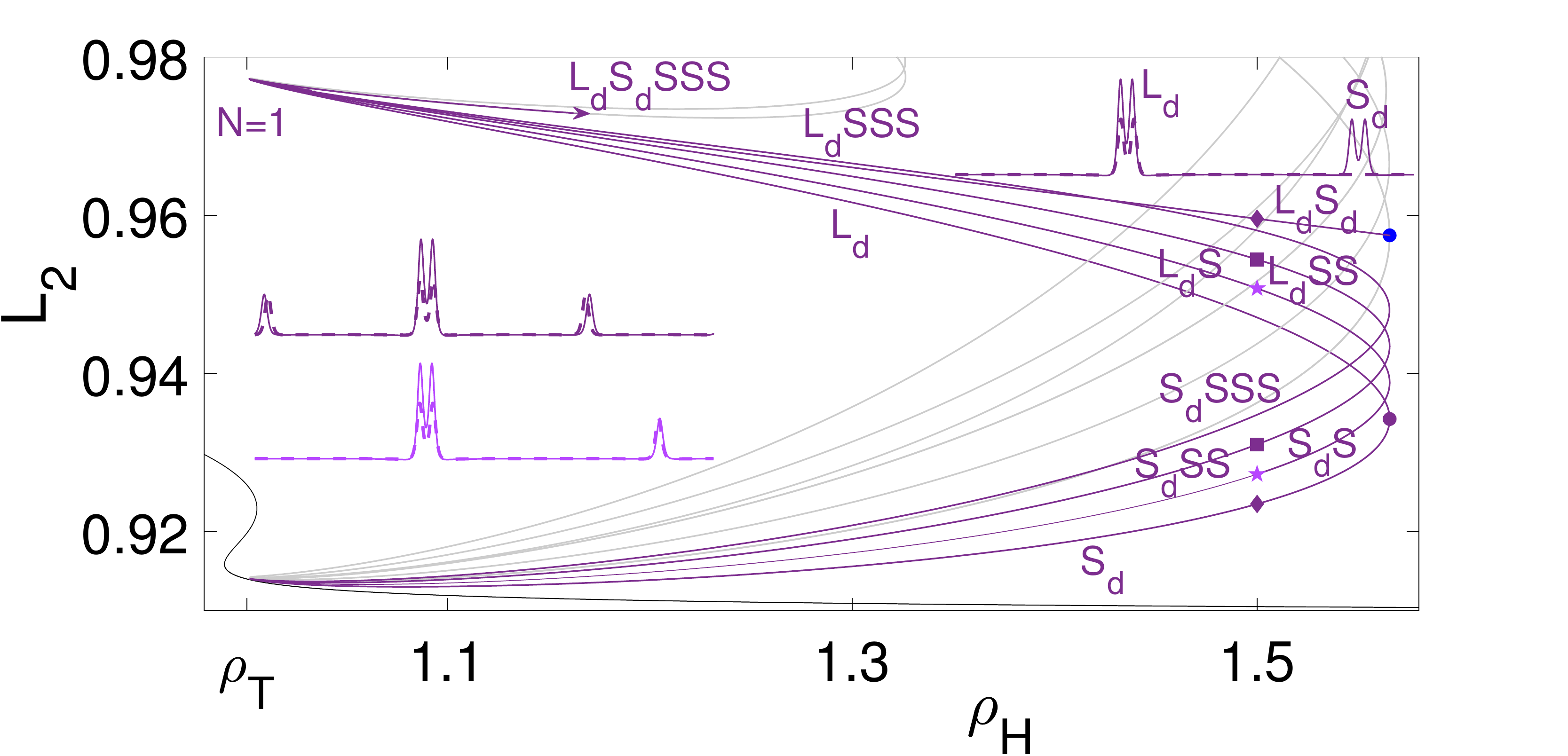}
		(b)\centering\includegraphics[width=0.75\textwidth]{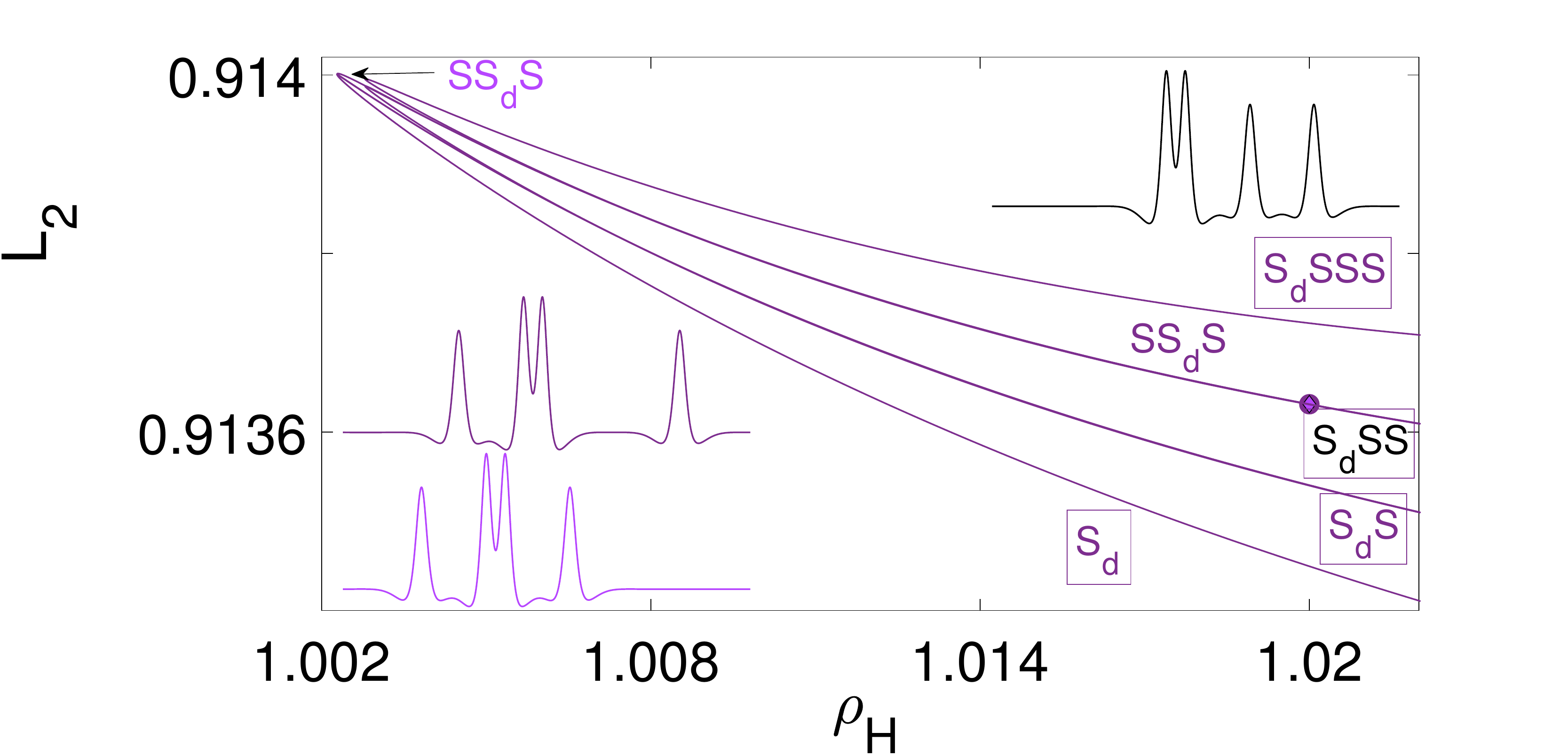}
		(c)\centering\includegraphics[width=0.75\textwidth]{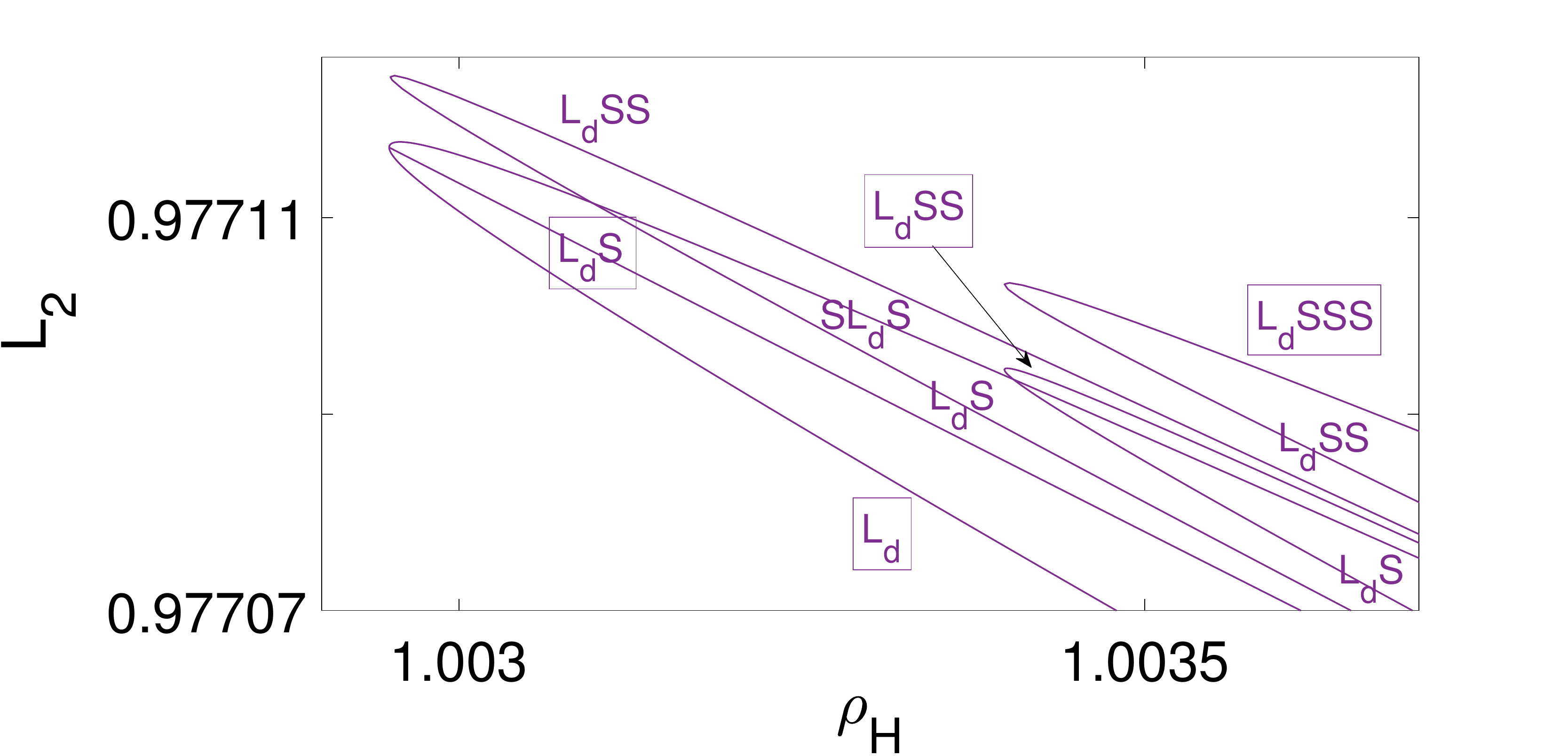}
		\caption{(a) Detail of the lower $N=1$ region of the bifurcation diagram in Fig.~\ref{fig:overall_split}. (b) Further detail of the region near the lower left folds. There are two folds and two distinct branches S$_d$SS and SS$_d$S corresponding to the profiles shown in the inset. (c) Further detail of the region near the upper left folds. There are four folds and three distinct branches L$_d$SS. The boxed labels indicate branches that extend to large values of $\rH$.}
		\label{fig:n1_split}
	\end{figure}

	Figure \ref{fig:n1_split} shows details of the $N=1$ state and associated branches. We refer to the 2-peak profiles on the $N=1$ branch using the labels S$_d$ and L$_d$ ($d$ for double-peak), depending on whether the profile lies below or above the $N=1$ fold. Thus S$_d$ turn into L$_d$ at the fold. The $N=1$ branch itself originates as S$_d$ close to $\rT$ and does so in a fold (Fig.~\ref{fig:n1_split}(b)). At the fold the branch turns, apparently smoothly, into an S$_d$SS state with the appearance of a pair of small single peaks; a third branch, of S$_d$S states bifurcates from the fold, just as in the case of the 1-peak states. The figure also shows a second fold, where a branch labeled SS$_d$S turns into S$_d$SSS through the addition of a single small peak. The S$_d$SS and SS$_d$S states are no longer the same, as can be seen from the profiles on each branch, computed at the same value of $\rH$, shown in the inset.
	
	Each of these five branches can be traced through its fold on the right (Fig.~\ref{fig:n1_split}(a)) and towards the associated upper left folds, a detail of which is shown in Fig.~\ref{fig:n1_split}(c) where analogous behavior takes place: L$_d$ turns, again apparently smoothly, into SL$_d$S, while an L$_d$S branch bifurcates directly from the fold. Two additional folds are also present, involving the transformation from L$_d$S to L$_d$SS and the transformation from a different L$_d$SS into L$_d$SSS. The boxed labels indicate branches that extend to large values of $\rH$ and so are created in the folds on the right. One of these is the L$_d$S$_d$ branch that terminates near the $N=2$ fold: as one follows this branch in the direction of increasing $\rH$ the height of S$_d$ increases and the separation of the two pulses adjusts in such a way that the state becomes an equispaced L$_d$L$_d$ state by the time the branch terminates.
	\begin{figure}[!t] %F13
		(a)\centering\includegraphics[width=0.75\textwidth]{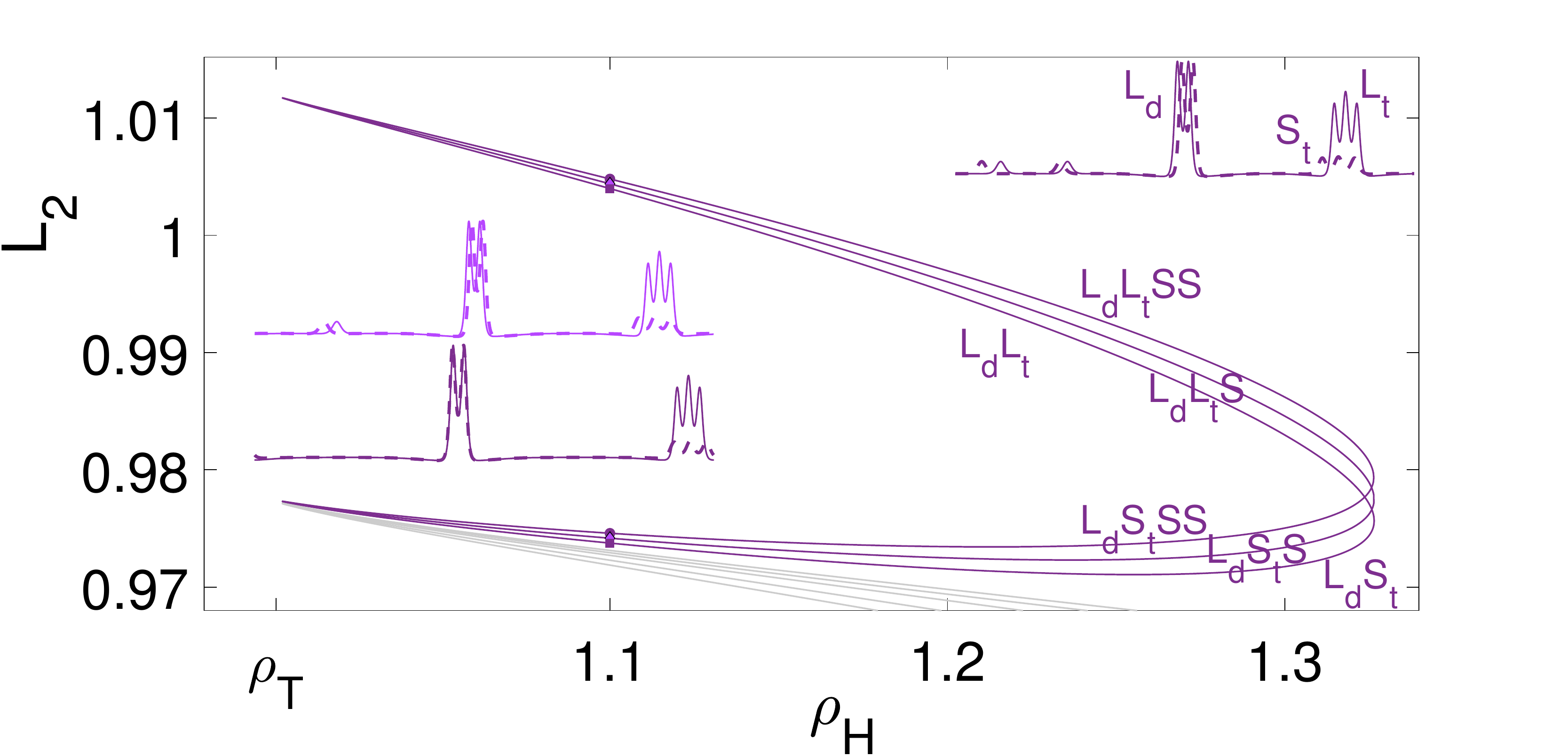}
		(b)\centering\includegraphics[width=0.75\textwidth]{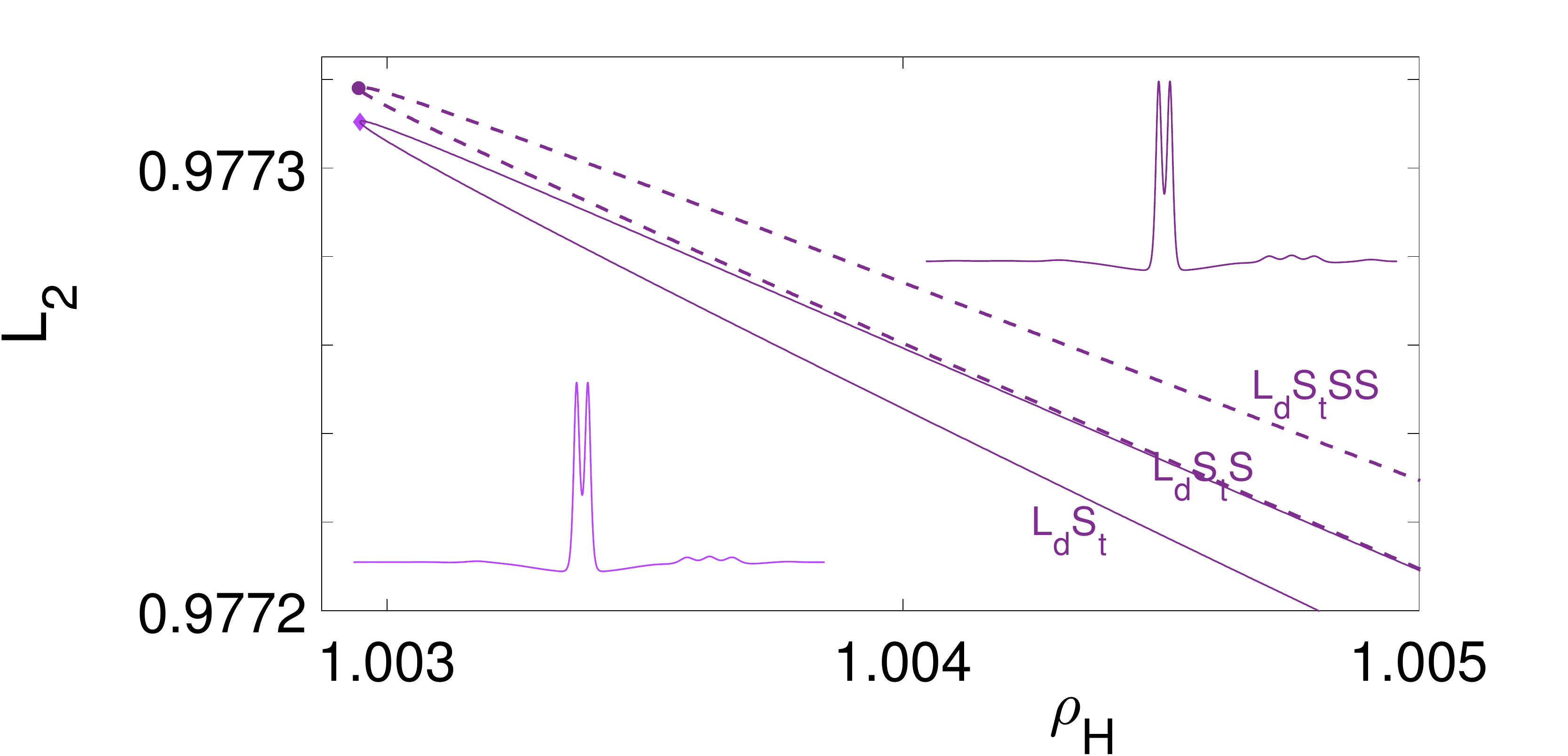}
		(c)\centering\includegraphics[width=0.75\textwidth]{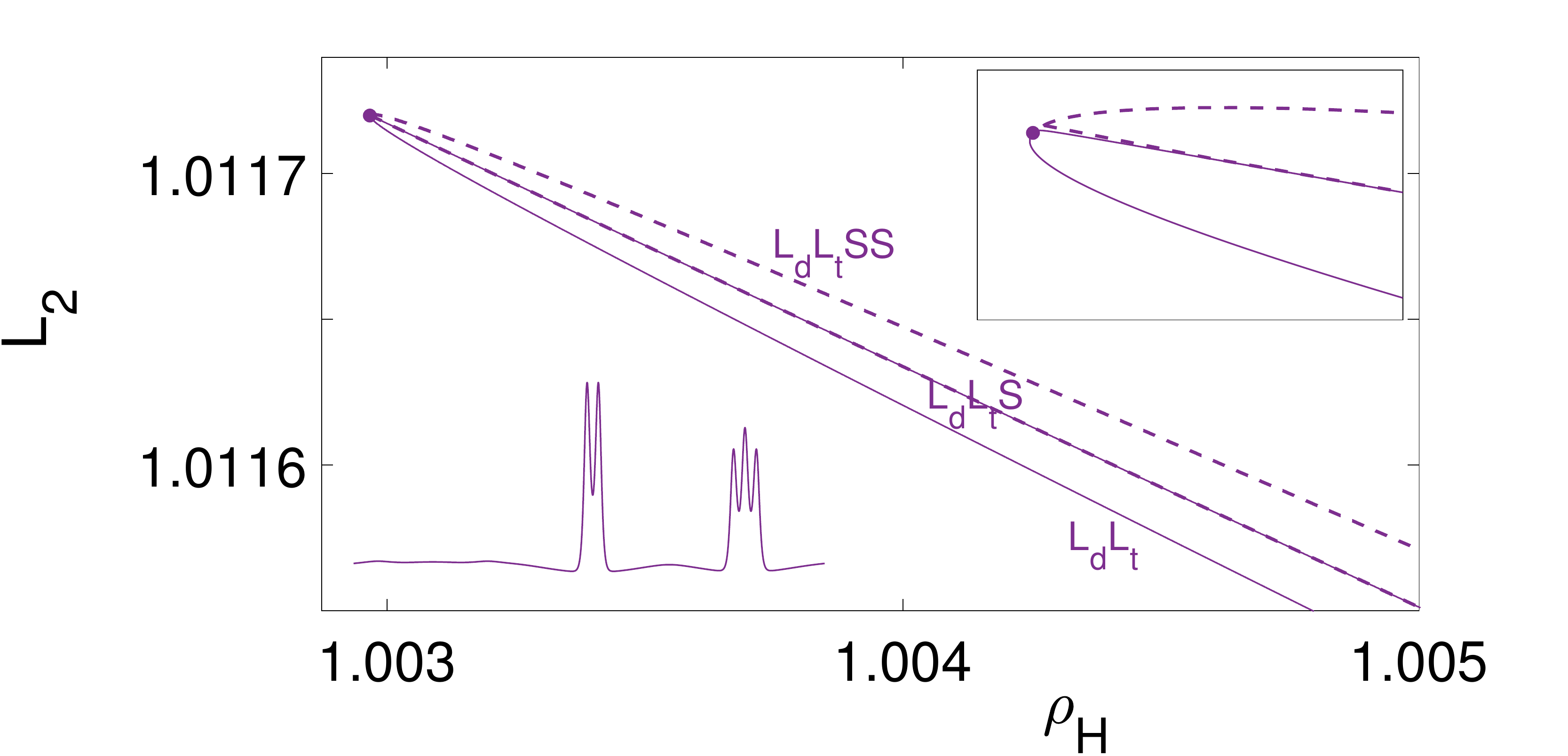}
		\caption{(a) Detail of the upper $N=1$ region of the bifurcation diagram in Fig.~\ref{fig:overall_split}. (b) Further detail of the region near the lower left folds. There are two folds and two distinct branches S$_d$SS and SS$_d$S corresponding to the profiles shown in the inset. (c) Further detail of the region near the upper left folds with two distinct branches L$_d$L$_t$S (see inset).}
		\label{fig:n2_split}
	\end{figure}
	
	Near the upper left folds of Fig.~\ref{fig:n1_split}(c) one also finds another set of folds, as shown in Fig.~\ref{fig:n2_split}(a). These branches correspond to bound states of a 2-peak state L$_d$ and a 3-peak state S$_t$ ($t$ for triple-peak) below the right fold and L$_t$ above it. As shown in Fig.~\ref{fig:n2_split}(b) these states are born in a manner that recapitulates the origin of the $N=1$ 2-peak states in Fig.~\ref{fig:n1_split}(b). The resulting states behave just like the $N=2$ 2-peak states: they extend to large $\rH$, pass through folds (Fig.~\ref{fig:n2_split}(a)) and then return to $\rH\simeq \rT$ with S$_t$ having grown into L$_t$ (Fig.~\ref{fig:n2_split}(c)). The main observation that merits a comment is that the right folds of these new 2-pulse states do not align with those of the original 2-pulse state associated with $N=1$. We believe this to be a finite size effect: if one removes the L$_d$ portion of the state, the 3-peak component S$_t$ behaves as an isolated 3-peak state on a much smaller domain, something like half the domain length $L=24$ (Fig.~\ref{fig:n2_split}(c)). Since the wavelength of the oscillations and hence the inter-peak separation is something like 2.8, the width of S$_t$ is of order 8.5 and hence quite comparable with the half-length of the domain. On such small effective domains the snaking of S$_t$ will depart from its asymptotic behavior on large domains, pushing the right folds inwards.
	
	Note that the 2-pulse states based on L$_d$ and S$_t$/L$_t$ form a completely distinct set of isolas from those based on L$_d$ and S$_d$/L$_d$. The latter do connect to the equispaced 2-pulse 2-peak states near the fold of the $N=2$ branch. The resulting $N=2$ states are in turn responsible for a similar sequence of 4-pulse states, shown in blue in Fig.~\ref{fig:overall_split}. These include the branch L$_d$SSS (Fig.~\ref{fig:n1_split}(a,c)). 
	%We expected an analogous branch, LSSS, to appear as part of the primary foliated snaking structure but, like LLLS, were unable to find it.
	
	It is clear that many more multipulse states of this type are present in the parameter range $\rT<\rH<\rSN$ but the above results suffice to illustrate the type of behavior expected. We mention that both multipulse and composite-peak states were computed in a four-dimensional non-Hamiltonian system by Champneys~\cite{champneys1994}, lending support to the conjecture at the end of the previous section.

	\section{Discussion and conclusions}
	
	Isolated, spatially localized peaks or spikes (spots in two dimensions) play an increasingly important role in many systems of physical, chemical, and biological interest. We have studied here in some detail one such system that exhibits a large multiplicity of different multipeak states and showed that these states are organized in a structure we call foliated snaking. We believe this structure to be universal and thus expect similar structures to be present in a variety of different systems. We have seen that in the present activator-inhibitor-substrate system, this structure is a associated with the presence of a subcritical Turing instability, and arises whenver the localized states familiar from other models exhibiting a subcritical pattern-forming instability such as the Swift-Hohenberg equation extend into a parameter region with real leading eigenvalues. In this case, the localized structures cannot lock via oscillatory tails and instead repel one another, forming a sea of isolated structures within the domain whose dynamical properties in time remain to be studied.
	
	A subcritical Turing instability is not the only one resulting in isolated localized structures. Earlier work~\cite{yochelis2008formation,parra2018bifurcation,ruiz,zelnik2017desertification} has shown that similar structures can be generated as a result of the presence of a fold or transcritical bifurcation of a homogeneous state, provided only that a Belyakov-Devaney (BD) point is present. The important feature of all these mechanisms is that they generate such states over a much larger range of parameter values than the better known homoclinic snaking mechanism. The latter generates related states but only within a typically narrow parameter range -- the snaking or pinning interval.
	
	In the present work, we studied in detail the origin of the primary $N$-peak states -- these form near $\rT$ as a consequence of a modulational instability of subcritical Turing states -- and their subsequent evolution with increasing $\rH$ via folds on the right at $\rSN$ through to their transformation into trains of unequal peaks near $\rT$ again. These states form the cross-links connecting the primary $N$-peak states between the left and right folds. Spatial resonances were found to play an essential role both in the origin of the primary $N$-peak states at small amplitude and in the generation of trains of unequal peaks near $\rSN$ and hence in the bifurcation structure associated with foliated snaking. Finally, the exchange point (EP) point was found to play the role usually associated with the BD point, and this role was found to be crucial for the transformation from the primary $N$-peak states to secondary states consisting of unequal peaks that form the cross-links in the bifurcation structure. We also made an extensive exploration of the so-called multipulse or subsidiary states that decorate the primary foliated snaking structures. These results only scratch the surface of the complexity exhibited by the present system.
	
	We have employed several different domain sizes $L$ to obtain these results. Normally the domain size $L$ is largely immaterial when one studies well-localized states. Here, however, the peaks in a multipeak state for $\rH>\rEP$ want to be apart as far as possible and hence the domain size sets the length scale for structures of this type, in contrast to the intrinsic wavelength $\ell=2\pi/|{\rm Im}(\lambda_{1,2})|$ that sets the length scale in the locking regime $\rT<\rH<\rEP$. Despite this inevitable dependence of our results on $L$ most aspects are essentially independent of it. This includes the locations $\rT$, $\rEP$ and $\rBD$ which are determined to high accuracy by the infinite domain calculation. In the nonlinear regime, the right folds accumulate on $\rSN$ as $N$ varies, a quantity that also almost independent of $L$ when $L$ is sufficiently large (Fig.~\ref{fig:rfolds}(b)). Since the transitions in the foliated snaking structure take place near $\rT$, $\rEP$ and $\rSN$ we conclude that this structure is almost independent of the domain size $L$ after all.
	\begin{figure}[!t] %F14
		\centering\includegraphics[width=0.8\textwidth]{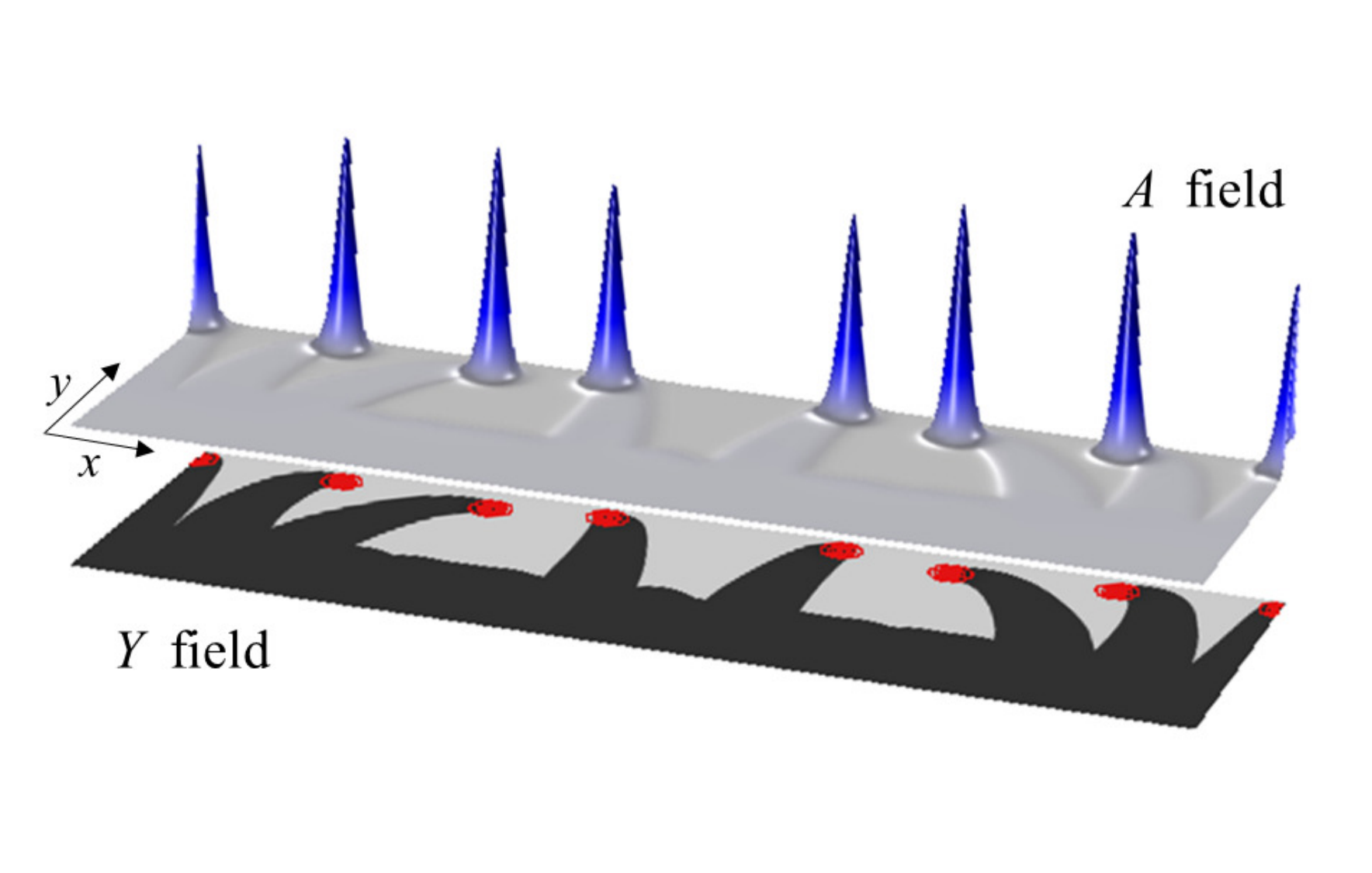}
		\caption{Representative snapshot of a solution obtained from a direct numerical simulations of~\eqref{eq:AI} as in Fig.~\ref{fig:intro} but at a later time.}
		\label{fig:disc}
	\end{figure}
	
	Temporal stability (determined via a standard eigenvalue formulation of the linearization of the system~\eqref{eq:AI}) shows that all stationary peaks and groups of peaks are linearly unstable in 1D while in 2D isolated spots in a background of $\vP_*$ and away from any front such as that in Fig.~\ref{fig:intro} may gain linear stability, facts that explain the oscillatory or decaying peak dynamics in 1D and persistence of peaks in 2D~\cite{yochelis2020nonlinear}. We are aware that in order to establish further significance of our results it is also necessary to investigate the temporal stability properties of the various states we have found, and this, in addition to the results of elaborate time-integration, will form the topic of a future contribution. Nevertheless, nucleation of side-branches can already be regarded as a nontrivial wavenumber selection problem because of the multiplicity of solutions that may give rise to distinct separation distances at which the peaks form, as shown in Fig.~\ref{fig:intro}. In practice, side-branches in pulmonary vascular patterns typically appear in the middle of the domain (far from gradients imposed by the leading peak)~\cite{yao2007matrix}, in between the initial branched point and the leading peak of the growing branch, and so the minimal length scale of the nucleation process depends also on the location of the right folds, i.e., for larger values of $\rH$ side-branches appear later and at large distances from the growing tip~\cite{yochelis2020nonlinear}. Moreover, the observed self-repulsion of the peaks explains the so-called \textit{avoidance} phenomenon, i.e., the absence of peak-peak collisions and the merging of counterpropagating peaks in 2D, as well as the potential inhibition of new peaks on nearby branches~\cite{hannezo2019multiscale}. Figure~\ref{fig:disc} shows that once the peaks feel a counterpropagating peak (perhaps reflected from the boundary due to Neumann boundary conditions in the $x$ direction) their trajectory is deflected in the $y$ direction, something that cannot happen in 1D.
	
	Evidently, the conjectured universal organization of peaks in the foliated snaking structure generated by a subcritical Turing instability provides a distinct mathematical description of homoclinic snaking in multivariable reaction--diffusion media. The results suggest that the nonlinear mechanisms behind both peak generation and wavenumber selection that appear to be involved in the side-branching model studied here lead to a robust phenomenon that is essential, among other examples, to studies of ecological systems, ranging from the initiation of root hairs~\cite{brena2014mathematical} to the generation of vegetation patches~\cite{meron2019vegetation} and the design of agroforestry systems~\cite{tzuk2020}.
	\\
	\\
	{\bf Acknowledgment}: We have benefited from discussions with Nicol\'as Verschueren. This work was supported in part by the National Science Foundation under grant DMS-1908891 (EK).

\end{document}